\documentclass[twocolumn]{aastex62}
\begin{document}
\title{Precursor Wave Emission Enhanced by Weibel Instability in
Relativistic Shocks} 

\author{Masanori Iwamoto} 
\affiliation{Department of Earth and Planetary Science, University of Tokyo,
7-3-1 Hongo, Bunkyo-ku, Tokyo 113-0033, Japan}

\author{Takanobu Amano} 
\affiliation{Department of Earth and Planetary Science, University of Tokyo,
7-3-1 Hongo, Bunkyo-ku, Tokyo 113-0033, Japan}

\author{Masahiro Hoshino} 
\affiliation{Department of Earth and Planetary Science, University of Tokyo,
7-3-1 Hongo, Bunkyo-ku, Tokyo 113-0033, Japan}

\author{Yosuke Matsumoto} 
\affiliation{Department of Physics, Chiba University, 1-33 Yayoi, Inage-ku,
Chiba, Chiba 263-8522, Japan}

\correspondingauthor{Masanori Iwamoto}
\email{iwamoto@eps.s.u-tokyo.ac.jp}

\begin{abstract}
 We investigated the precursor wave emission efficiency in magnetized
 purely perpendicular relativistic shocks in pair plasmas. We extended our
 previous study to include the dependence of upstream magnetic field
 orientations. We performed two-dimensional particle-in-cell simulations
 and focused on two magnetic field orientations: the magnetic field to
 be in the simulation plane (i.e., in-plane configuration) and
 perpendicular to the simulation plane (i.e., out-of-plane
 configuration). Our simulations in the in-plane configuration
 demonstrated that not only extraordinary but also ordinary mode waves
 are excited. We quantified the emission efficiency as a function of the
 magnetization parameter $\sigma_e$ and found that the large-amplitude
 precursor waves are emitted for a wide range of $\sigma_e$. We found
 that especially at low $\sigma_e$, the magnetic field generated by
 Weibel instability amplifies the ordinary mode wave power. The
 amplitude is large enough to perturb the 
 upstream plasma, and transverse density filaments are generated as in
 the case of the out-of-plane configuration investigated in the previous
 study. We confirmed that our previous conclusion holds regardless of
 upstream magnetic field orientations with respect to the two-dimensional
 simulation plane. We discuss the precursor wave emission in
 three dimensions and the feasibility of wakefield acceleration in
 relativistic shocks based on our results. 
 
\end{abstract}


 \section{Introduction}\label{sec:intro}

 Observations of active galactic nuclei (AGNs) and gamma ray bursts
 (GRBs) usually show broad nonthermal spectra
 \citep[e.g.,][]{Kaneko2006, Abdo2010}, which are believed to
 originate from synchrotron radiation and inverse Compton scattering of
 relativistic electrons. Since the relativistic outflow from the central
 compact object is the common feature in AGNs and GRBs
 \citep[e.g.,][]{Gehrels2009, Lister2016}, relativistic shocks can be
 formed upon interaction between the jets and the interstellar
 medium. The relativistic shocks are assumed to play an important role
 for generating such nonthermal electrons. 
 
 Previous one-dimensional (1D) particle-in-cell (PIC) simulations showed
 that synchrotron maser instability (SMI) is the significant dissipation
 mechanism for relativistic magnetized shocks
 \citep[e.g.,][]{Langdon1988, Gallant1992, Hoshino1992, Amato2006}. The
 SMI is driven by particles reflected off the shock-compressed magnetic 
 field in the shock-transition region and emits electromagnetic waves
 of extraordinary mode (X-mode) both upstream and downstream
 \citep{Hoshino1991}. Since the electromagnetic precursor waves have a
 non-negligible fraction of the upstream kinetic 
 energy, the upstream flow is significantly perturbed by the precursor
 wave \citep{Lyubarsky2006}. \cite{Hoshino2008} demonstrated that the
 wave power is strong enough to induce wakefield in the upstream and
 that nonthermal electrons are generated by wakefield acceleration
 \citep[WFA;][]{Tajima1979, Chen2002} in 1D relativistic shocks
 propagating in magnetized ion--electron plasmas. 

 In multidimensional systems, it is well known that Weibel instability
 \citep[WI;][]{Weibel1959, Fried1959} develops in the transition
 region of weakly magnetized shocks. \added{The WI is widely studied in laser
 plasma as well as astrophysics 
 \cite[e.g.,][]{DAngelo2015, Huntington2015, Huntington2017, Park2015}.} 
 Previous PIC simulation studies in multiple dimensions \replaced{indeed
 showed that the shock transition is dominated by the WI }{demonstrated that
 the WI grows into substantial amplitude in the shock-transition region}
 at low magnetization 
 $\sigma_e = \omega_{ce}^2/\omega_{pe}^2 \lesssim 10^{-2}$
 \citep[e.g.,][]{Spitkovsky2005,Sironi2013}. Here, $\omega_{ce}$ is  the
 relativistic electron cyclotron frequency and $\omega_{pe}$ is the
 proper electron plasma frequency. \replaced{The effective temperature 
 anisotropy in the shock-transition region induced by reflected
 particles drives the WI (e.g., Kato 2017; Chang et
 al. 2008). The maximum growth rate of the WI including 
 relativistic effects scales as $\omega_{pe}$ for sufficiently strong
 anisotropy (see, e.g., Yang et al. 1993; Achterberg et al. 2007;
 Schaefer-Rolffs \& Tautz 2008). In 
 contrast, the growth rate of the SMI is on the order of $\omega_{ce}$
 (Hoshino 1991).}{The effective temperature 
 anisotropy induced by reflected particles in the shock-transition
 region provides the free energy source for the development of the WI
 \citep[e.g.,][]{Kato2007, Chang2008}. The linear theory including 
 relativistic effects 
 showed that the maximum growth rate of the WI is on the order of
 $\omega_{pe}$ \citep[see, e.g., ][]{Yang1993, Achterberg2007,
 Schaefer2008}, whereas that of the SMI is on the order of $\omega_{ce}$ 
 \citep{Hoshino1991}.}
 Since both instabilities are excited from the
 same free energy source in the same 
 region and $\omega_{pe}$ is much greater than $\omega_{ce}$ for
 $\sigma_e \ll 1$, it was believed that the WI dominates 
 over the SMI and the precursor wave emission could be shut off in
 multidimensional shocks.

 Recently, by using two-dimensional (2D) PIC simulations, we have shown
 that the SMI can coexist with the WI and that the precursor wave
 emission continues to persist even in the Weibel-dominated 
 regime \citep{Iwamoto2017}. We also showed that the
 wave power is sufficient enough to induce wakefield for a wide range of
 magnetization parameter $\sigma_e$. Based on the results, we suggested
 that external shocks 
 in the relativistic jets from GRBs may be important sites for the
 production of ultra-high-energy cosmic rays via WFA.

 However, in the previous work, we focused only on
 perpendicular shock with the upstream ambient magnetic field
 perpendicular to the simulation plane (i.e., out-of-plane
 configuration). One may also choose the upstream ambient 
 magnetic field to be in the simulation plane (i.e., in-plane
 configuration), which may in general change the shock dissipation
 physics because the degree of freedom in this case becomes three rather
 than two in the out-of-plane configuration \citep[e.g.,][]{Amano2009}.
 In fact, \cite{Sironi2013} reported that the 
 particle acceleration efficiency in 2D perpendicular shocks depends on
 the orientation of the pre-shock magnetic field. Therefore, in this
 study, we consider the in-plane configuration and investigate the
 physics of magnetized perpendicular shocks, especially the
 electromagnetic wave emission by the SMI. We quantify the precursor
 wave emission efficiency and discuss the effects of the magnetic
 field configuration on the WFA combining this study with our previous
 results.

 This paper is organized as follows. First, Section \ref{sec:setup}
 describes our simulation setup. In Section \ref{sec:shock}, we show the
 global structure of relativistic magnetized shocks for relatively high
 and low magnetization, respectively. In Section \ref{sec:precursor},
 the properties of precursor waves are analyzed. In
 Section \ref{sec:discussion}, we discuss the wave excitation mechanism
 and the feasibility of the WFA in relativistic magnetized
 shocks. Finally, Section \ref{sec:summary} summarizes this study.

 \section{Simulation setup}\label{sec:setup}

\begin{figure}[htb!]
  \plotone{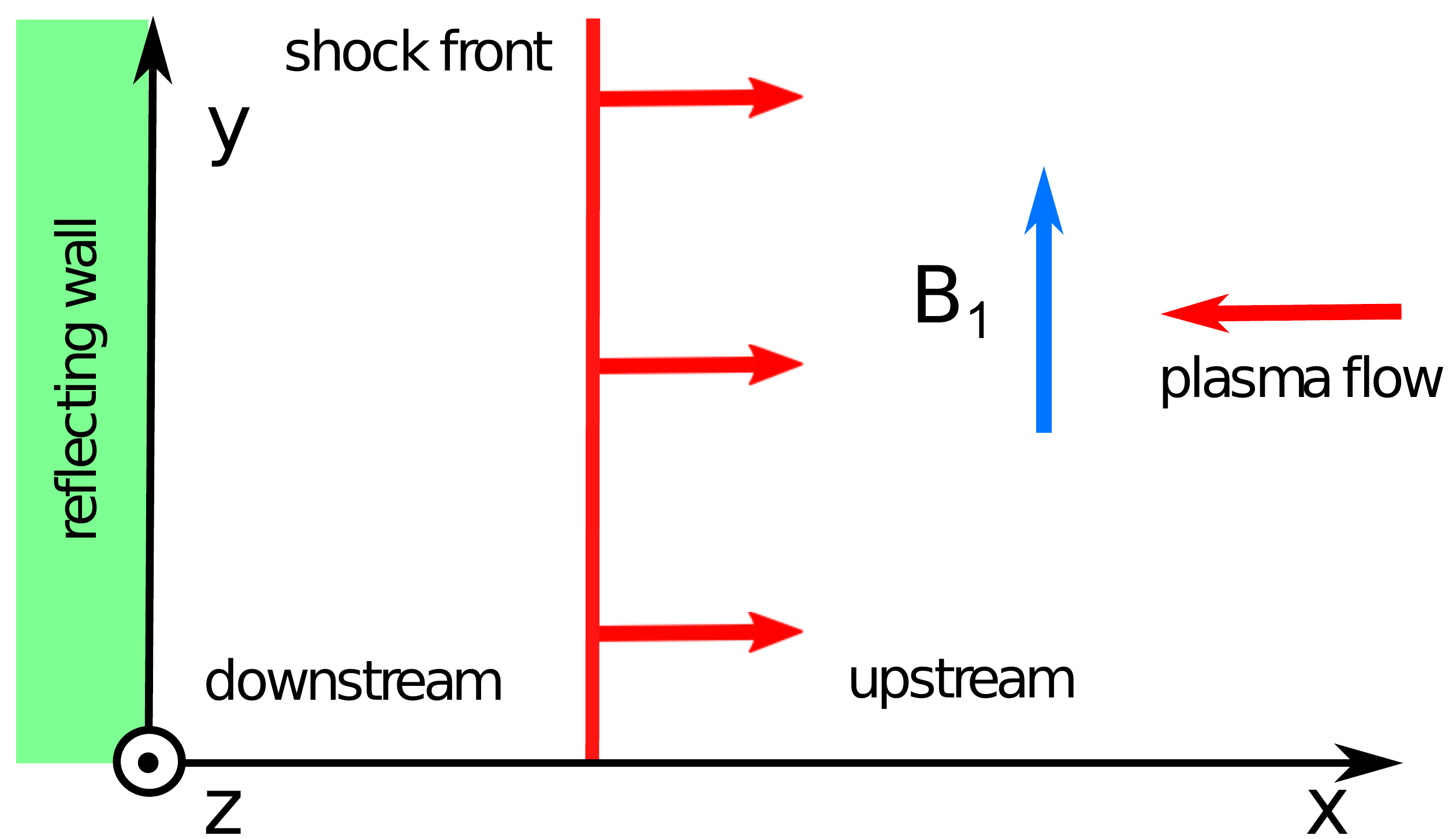}
  \caption{\replaced{Coordinate system and geometry of the simulation. The
  upstream ambient magnetic field $B_1$ is
  aligned in the $y$ direction (cf. Iwamoto et al. 2017).}{Coordinate
  system and the orientation of the upstream ambient magnetic field in
  the present simulation \citep[cf.][]{Iwamoto2017}.}}
  \label{configuration}
 \end{figure}

 We carried out simulations of 2D perpendicular shocks in
 electron--positron plasmas using an electromagnetic PIC code
 \citep{Matsumoto2013, Matsumoto2015}. The basic 
 configuration of our simulations is almost identical to our previous
 simulation \citep{Iwamoto2017} and schematically illustrated in Figure
 \ref{configuration}. We changed only the direction of the upstream
 ambient magnetic field $B_1$ from the out-of-plane direction ($z$
 direction in our coordinate system) to the in-plane direction ($y$
 direction).

 Our simulation domain is in the $x$--$y$ plane with periodic boundary
 conditions in the $y$ direction and the number of grids in each
 direction is $N_x \times N_y = 20,000 \times 1,680$. A cold pair stream
 is continuously injected along $-x$ direction with a bulk Lorentz factor
 $\gamma_1=40$ from the right-hand boundary and elastically reflected at
 the left-hand boundary.  The shock wave is excited by the interaction
 between the returning particles and the incoming plasma flow, and
 propagates toward $+x$ direction. The number of particles per cell in
 the upstream is $N_1 \Delta x^2 = 64$ for both electrons and positrons,
 where $\Delta x$ is the grid size. The grid size is fixed to
 $\Delta x/(c/\omega_{pe})= 1/40$ throughout in this study, where $c$ is
 the speed of light and the $\omega_{pe}$ is the proper electron plasma
 frequency. The proper electron plasma frequency is defined as follows: 
 \begin{equation}
  \label{eq:wpe}
  \omega_{pe}=\sqrt{\frac{4 \pi N_1 e^2}{\gamma_1 m_e}}.
 \end{equation}  
 The number of particles per cell and the grid size are motivated by the
 numerical convergence study of 1D simulations
 \citep[see][Appendix A]{Iwamoto2017}.
 The time step is set to be $\omega_{pe} \Delta t = 1/40$ in order to
 minimize the effect of the numerical Cherenkov instability
 \citep{Ikeya2015}. For more details, please refer our previous paper
 \citep{Iwamoto2017}. 
  
 As in our previous study, we investigated the dependence of the
 precursor wave emission on the magnetization parameter $\sigma_e$:
 \begin{equation}
  \sigma_e = \frac{B_1^2}{4 \pi \gamma_1 N_1 m_e c^2} =
   \frac{\omega_{ce}^2}{\omega_{pe}^2},  
 \end{equation}
 where $\omega_{ce}$ is the relativistic electron cyclotron frequency:
 \begin{equation}
  \omega_{ce} = \frac{eB_1}{\gamma_1m_ec}.
 \end{equation}
 More specifically, we discuss the results obtained from the following
 eight runs: $\sigma_e = 1$, $3 \times 10^{-1}$, $1 \times 10^{-1}$, $3\times
 10^{-2}$, $1\times 10^{-2}$, $3\times 10^{-3}$, $1\times 10^{-3}$ and
 $3\times 10^{-4}$.

 \section{Global Shock Structure}\label{sec:shock}
  \subsection{High-$\sigma_e$ Case}\label{subsec:highsig}

\begin{figure*}[htb!]
   \plotone{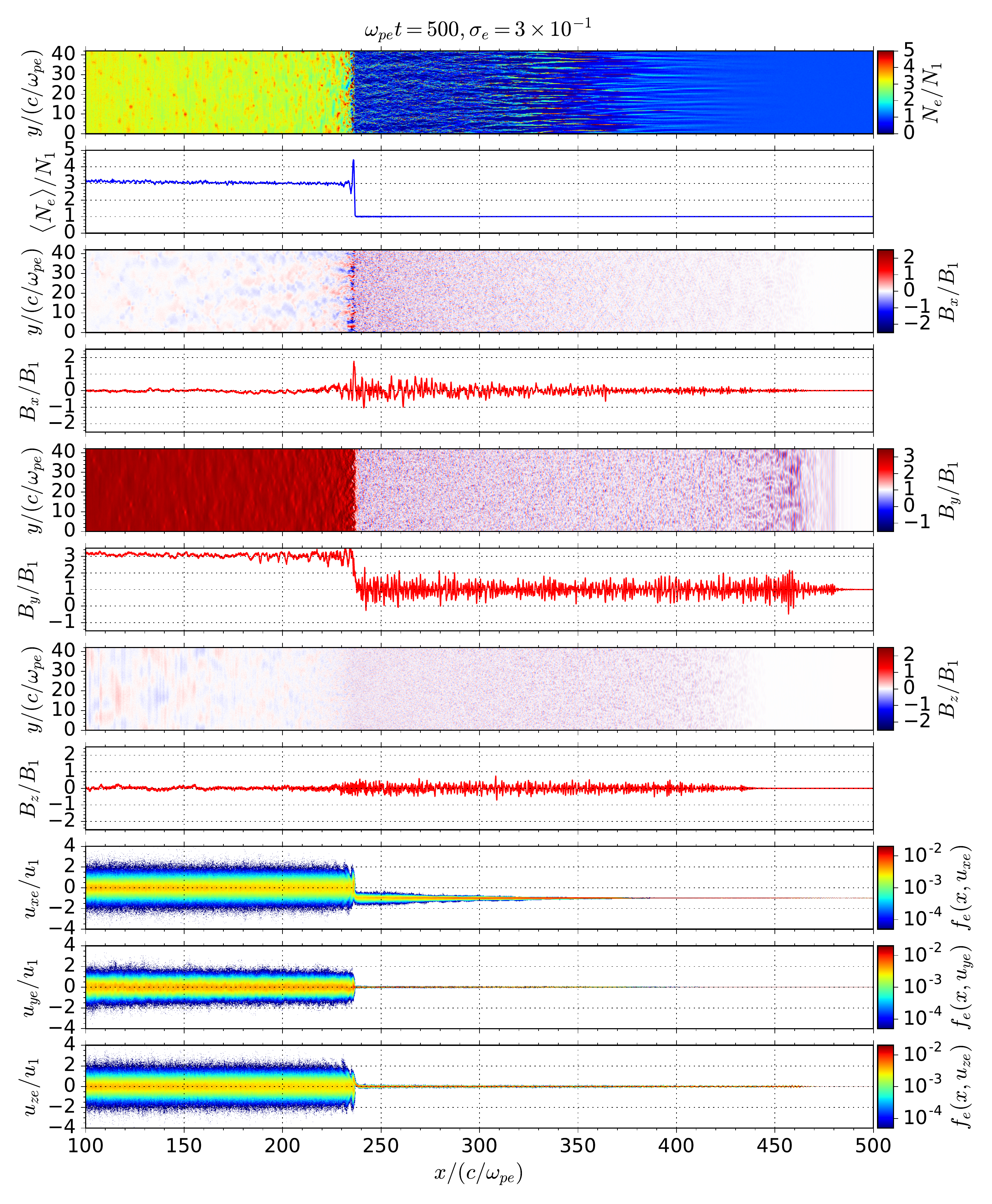}
   \caption{\replaced{Global shock structures at $\omega_{pe}t=500$ for $\sigma_e
   = 3 \times 10^{-1}$. From top to bottom, the electron number density 
   $N_e$, the transversely averaged electron number density $\langle N_e
   \rangle$, the $x$ component of the magnetic field $B_x$, the 1D profile
   for $B_x$ taken along $y/(c/\omega_{pe}) = 21$, the in-plane magnetic
   field $B_y$, the 1D profile for $B_y$ taken along
   $y/(c/\omega_{pe}) = 21$, the out-of-plane magnetic field $B_z$, the
   1D profile for $B_z$ taken along $y/(c/\omega_{pe}) = 21$ and the
   phase-space plots of electrons in the $x$--$u_{xe}$, $x$--$u_{ye}$
   and $x$--$u_{ze}$ planes are shown.}{Shock structure and electron
   phase space from a
   simulation with $\sigma_e = 3 \times 10^{-1}$ at
   $\omega_{pe}t=500$. From top to bottom, the electron number density 
   $N_e$, the average density $\langle N_e \rangle$, the
   longitudinal magnetic field $B_x$, 1D cut for $B_x$ along
   $y/(c/\omega_{pe}) = 21$, 
   the in-plane magnetic field $B_y$, 1D cut for $B_y$ along
   $y/(c/\omega_{pe}) = 21$, the out-of-plane magnetic field $B_z$, 1D
   cut for $B_z$  along 
   $y/(c/\omega_{pe}) = 21$ and the electron
   phase-space plots of $x$--$u_{xe}$, $x$--$u_{ye}$ and $x$--$u_{ze}$
   are shown.}}    
   \label{in-plane_high}
  \end{figure*}

  First, we discuss the overview of the global shock structure for
  relatively high $\sigma_e$. Figure \ref{in-plane_high} is the global
  shock structure at $\omega_{pe} t = 500$ for $\sigma_e = 3 \times
  10^{-1}$. The electron number density $N_e$, the electron number
  density averaged along the $y$ axis $\langle N_e \rangle$, the $x$
  component of the magnetic field $B_x$,  1D cut along $y =
  21c/\omega_{pe}$ for $B_x$, the in-plane magnetic field $B_y$, 1D cut
  along $y = 21c/\omega_{pe}$ for $B_y$, the out-of-plane magnetic field
  $B_z$, 1D cut along $y/(c/\omega_{pe}) = 21$ for $B_z$ and the electron 
  phase-space density $x$--$u_{xe}$, $x$--$u_{ye}$ and $x$--$u_{ze}$
  integrated over the $y$ direction are shown from top to bottom. All
  quantities are normalized by the corresponding upstream values. Note
  that our 2D simulations track all three components of the particle
  velocity and electromagnetic field. A well-developed shock structure
  is formed at this time, and the shock front is clearly seen at
  $x/(c/\omega_{pe}) \sim 235$.

  At the shock front, fluctuations in $B_x$ are generated. We think
  the magnetic field fluctuations may be attributed to instabilities
  excited in the shock-transition region. One of the possible
  instabilities for this case is the Alfv\'en-ion-cyclotron instability,
  which is an electromagnetic instability on the Alfv\'en mode branch
  driven by a temperature anisotropy \citep[e.g.,][]{Winske1988}. 
  We perform linear analysis for a relativistic pair plasma with a cold
  ring distribution and indeed find a similar instability. This
  instability may be the cause of fluctuations in $B_x$ and the magnetic
  field energy is eventually amplified up to $10\%$--$20\%$ of the
  upstream kinetic energy. Although fluctuations in $B_z$ at the shock
  front may also be generated by the instability, the fluctuations
  start decreasing in time after $\omega_{pe}t \sim 140$ and are not
  clearly seen at this time.  
  
  The wave magnetic fields $\delta B_y$ are visible in the upstream
  region. The electromagnetic waves are continuously emitted from the
  shock front and persist with large amplitude. Remember that the
  upstream ambient magnetic field is in the $y$ direction. The wave
  magnetic field is polarized in the $y$ direction 
  and parallel to the ambient magnetic field, which is the signature of the
  X-mode wave (see Section \ref{subsec:mode}). This result is consistent
  with both the linear theory \citep{Hoshino1991} and the previous 2D
  simulation \citep{Iwamoto2017}. The oblique propagation of these
  X-mode waves may be responsible for the $x$ component of the
  fluctuating magnetic field $\delta B_x$ in the upstream region. Since
  $\delta B_x$ is very small compared to $\delta B_y$, we mainly 
  consider $\delta B_y$ in our analysis. We think that the waves in the
  region $x/(c/\omega_{pe}) \gtrsim 460$ are contaminated by the
  initial and boundary conditions. Therefore, we excluded this region
  from our analysis presented below.
  
  The wave magnetic fields $\delta B_z$ are identified in the upstream
  region. They also appear to be electromagnetic precursor waves emitted
  from the shock front. The wave magnetic field is polarized in the $z$  
  direction, and thus the wave mode is the ordinary mode (O-mode; see
  Section \ref{subsec:mode}). This is unexpected because the
  linear theory of the SMI showed that the growth rate of the O-mode is
  finite at oblique propagation but much smaller than that of the X-mode
  \citep[see, e.g.,][]{Wu1979, Lee1980, Melrose1982, Melrose1984}.
  The amplitude of the O-mode wave is smaller than that of the X-mode
  wave but non-negligible. The tip of the O-mode wave is behind that of the 
  X-mode wave. This delay should result from the difference of the
  generation time since both have group velocities almost equal to the
  speed of light. The X-mode waves are generated by the SMI soon after the  
  shock formation in the initial phase of the simulation. In contrast,
  the generation of the O-mode waves seems to become effective after
  $\omega_{pe}t \sim 80$, which is estimated from the time evolution of
  the wave magnetic field $\delta B_z$. We discuss how the O-mode waves
  are excited in Section \ref{subsec:omode} for details.

  As in the case of our previous simulation, transverse density
  filaments are formed in the upstream region. This again indicates that
  the precursor waves remain large amplitude and coherent in 2D systems.
  
  \subsection{Low-$\sigma_e$ Case}\label{subsec:lowsig}

\begin{figure*}[htb!]
   \plotone{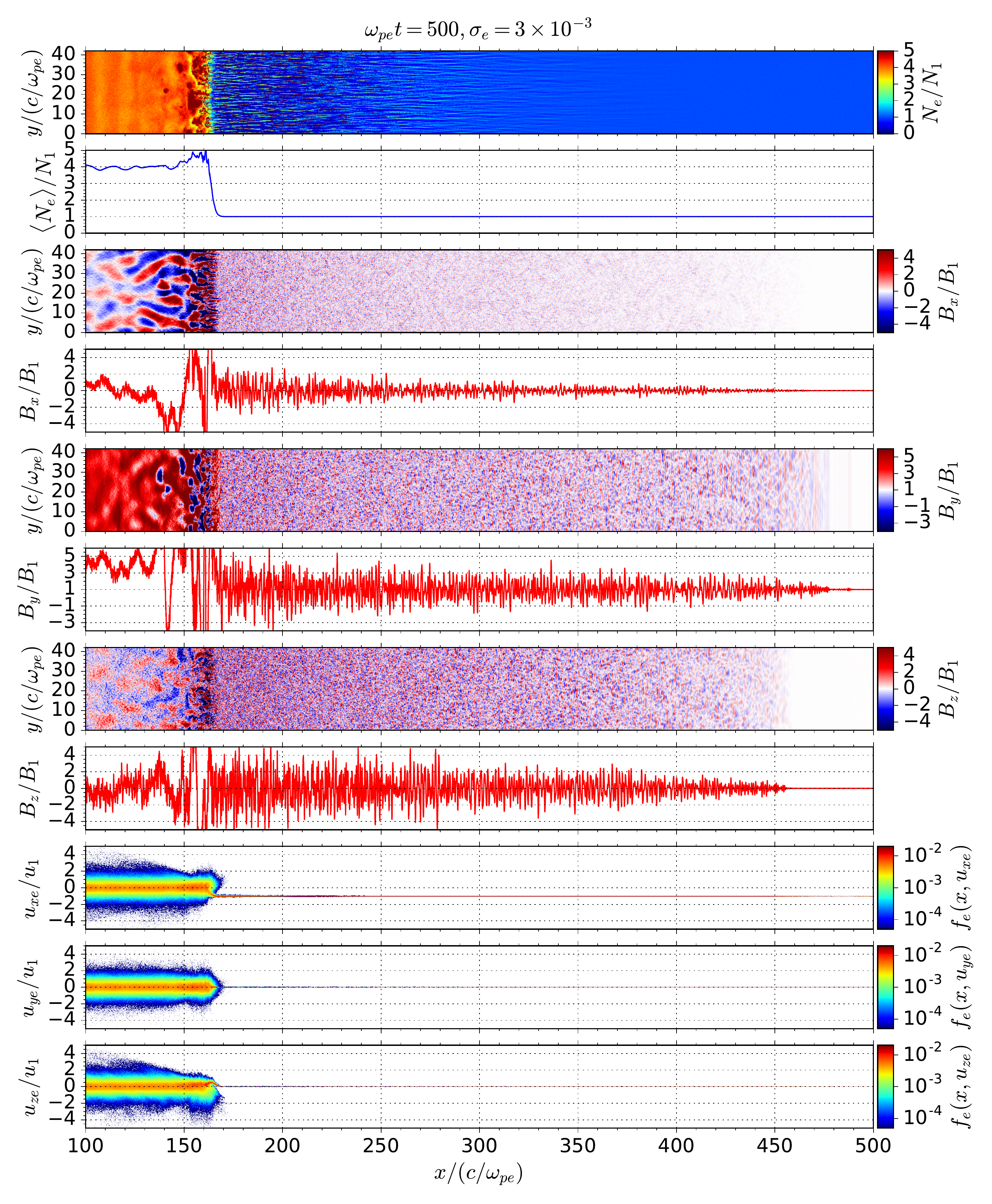}
   \caption{\replaced{Global shock structures at $\omega_{pe}t=500$ for
   $\sigma_e= 3 \times 10^{-3}$. See the caption of Figure 2 for
   details.}{Shock structure and electron phase 
   space from a simulation with $\sigma_e = 3 \times 10^{-3}$ at
   $\omega_{pe}t=500$. The format is the same as Figure
   \ref{in-plane_high}.}}
   \label{in-plane_low}
  \end{figure*}

  Here we discuss the overall shock structure for relatively
  low $\sigma_e$. Figure \ref{in-plane_low} is the global shock structure
  at $\omega_{pe} t = 500$ for $\sigma_e = 3 \times 10^{-3}$. The format
  is the same as Figure \ref{in-plane_high}. A well-developed shock
  front is distinctly visible at $x/(c/\omega_{pe}) \sim 160$.

  The filamentary magnetic field, expected for the structure of the
  Weibel-generated magnetic field, is seen at the shock front in the $x$
  and $z$ direction. In the in-plane configuration, previous works
  indeed showed that the WI excites $B_x$ as well as $B_z$
  \citep[e.g.,][]{Matsukiyo2006}. Our linear analysis arrives
  at the same conclusion (see Appendix \ref{sec:linear}). Thus we think
  that the fluctuations in $B_x$ and $B_z$ near the shock front is
  attributed to the WI. The maximum magnetic field energies for both
  components reach about $10\%$--$20\%$ of the upstream kinetic energy,
  which is consistent with the previous studies 
  \citep{Kato2007, Chang2008, Sironi2011}. However, the length of the
  Weibel filaments are shorter than that in the out-of-plane
  configuration \citep[see][Figure 3]{Iwamoto2017}. As we already
  explained, the WI is driven unstable by the effective temperature
  anisotropy induced by the reflected particles 
  \citep{Kato2007, Chang2008}. Therefore, the difference of the Weibel
  filaments originate from the relatively short reflected particle beam
  in the in-plane configuration. 

  One of the possible causes for the relatively short reflected particle
  beam is the relatively strong
  shock-compressed magnetic field in the in-plane 
  configuration. Recall that the degree of freedom is three in the   
  in-plane configuration. The adiabatic index for a relativistic ideal
  gas is $4/3$ rather than $3/2$ in the in-plane configuration and thus the
  compression ratio is greater than that in the out-of-plane
  configuration. Since particles are reflected off the shock-compressed
  magnetic field, the strong magnetic field compared to the out-of-plane
  case may result in a shorter length for the reflected particle beam.
  
  The suppression of the cross-field diffusion in the in-plane
  configuration may also contribute to the relatively short reflected
  particle beam. \cite{Jokipii1993} and \cite{Jones1998} mathematically
  proved that charged particles cannot move further than one Larmor
  radius from a given magnetic field if there are one or more ignorable
  coordinates. A notable exception is a 
  2D system with the out-of-plane magnetic field, which thus allows
  particles to diffuse across the magnetic field. In contrast, the
  diffusion of particles back into the upstream is prohibited in the
  in-plane configuration. Therefore, the length of the reflected
  particle beam in the in-plane case may become shorter than that in the
  out-of-plane case.

  The precursor waves are observed both in $B_y$ and $B_z$ and the delay
  of the O-mode precursor wave is identified in this case
  as well. The generation time of the O-mode wave may be estimated to be
  $\omega_{pe}t \sim 40$. The amplitude of the 
  O-mode wave is comparable to that of the X-mode wave unlike the
  high-$\sigma_e$ case. The $\sigma_e$ dependence is discussed in Section 
  \ref{subsec:dependence} in more detail. Notice that clear density
  filaments are observed in the precursor region in this case as well. 
  
  \section{Precursor Wave}\label{sec:precursor}
  \subsection{Wave Mode}\label{subsec:mode}
  
  As we mentioned in Section \ref{sec:shock}, the X-mode and O-mode
  electromagnetic waves are observed in the in-plane configuration.
  \replaced{The X-mode wave has a fluctuating component of the magnetic
  field $\delta B_X$ parallel to the ambient magnetic field $B_1$
  ($\delta B_X \parallel B_1$) and perpendicular to the wavenumber
  vector $k_X$ ($\delta B_X \perp k_X$). In a pair plasma, the X-mode
  wave is linearly polarized because the wave electric field 
  $\delta E_X$ is perpendicular to the ambient magnetic field
  ($\delta E_X \perp B_1$) and the wave vector 
  ($\delta E_X \perp k_X$). On the other hand, the fluctuating magnetic
  field of the O-mode wave $\delta B_O$ is 
  perpendicular to the ambient magnetic field $B_1$ ($\delta B_O \perp B_1$)
  and the wavenumber vector $k_O$ ($\delta B_O \perp k_O$). The fluctuating
  electric field $\delta E_O$ is parallel to the ambient magnetic field
  ($\delta E_O \parallel B_1$) and perpendicular to the wavenumber
  vector ($\delta E_O \perp k_O$), and thus the O-mode wave is linearly
  polarized as well.}{Both of the waves propagate perpendicular to the
  ambient magnetic field and are linearly polarized in pair plasmas. The
  wave magnetic field of the X-mode is parallel to the ambient
  magnetic field, whereas that of the O-mode is perpendicular.}  

\begin{figure*}[htb!]
   \plottwo{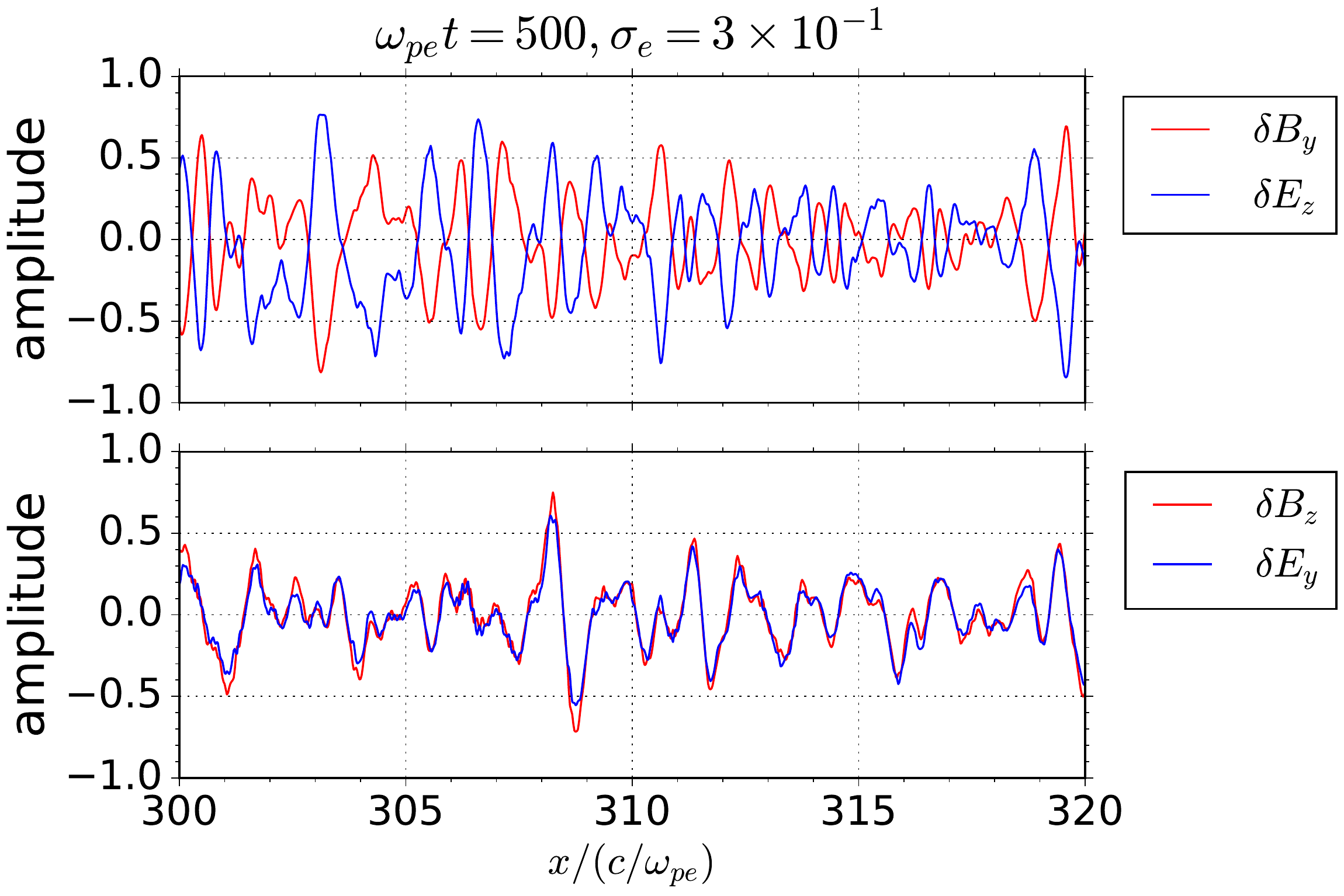}{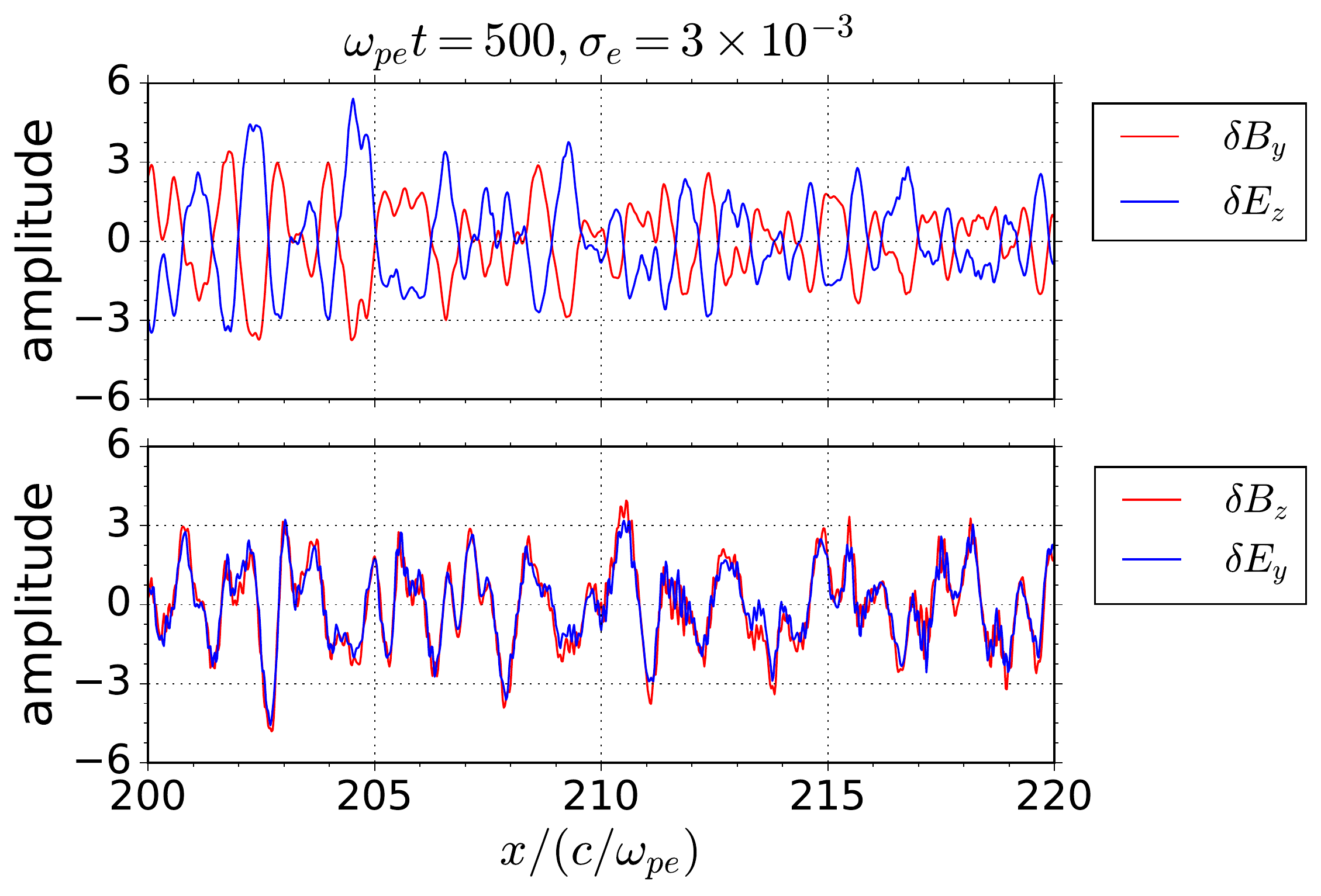}
   \caption{$y$ and $z$ components of wave electromagnetic fields at
   $\omega_{pe}t =500$ for $\sigma_e = 3 \times 10^{-1}$ (left) and
   $\sigma_e = 3 \times 10^{-3}$ (right). The electromagnetic fields are
   normalized by the upstream ambient magnetic field. The red and blue
   solid lines indicate the magnetic field and electric field,
   respectively.}
   \label{mode}
  \end{figure*}

  Figure \ref{mode} is the enlarged view of the region in
  $300 \le x/(c/\omega_{pe}) \le 320$ for $\sigma_e = 3\times 10^{-1}$
  (left) and $200 \le x/(c/\omega_{pe}) \le 220$ for
  $\sigma_e = 3 \times 10^{-3}$ (right), and shows the $y$ and $z$
  components of the wave electromagnetic fields at
  $\omega_{pe}t=500$. The electromagnetic 
  fields are normalized by the upstream ambient magnetic field
  $B_1$. The top panels show the $y$ component of the wave magnetic
  field $\delta B_y$ and the $z$ component of the wave electric field
  $\delta E_z$, and the bottom panels show the $z$ component of the wave
  magnetic field $\delta B_z$ and the $y$ component of the wave electric
  field $\delta E_y$. The red and blue lines indicate the magnetic field
  and electric field, respectively. Recall that the upstream ambient
  magnetic field $B_1$ is oriented along the $y$ axis. The
  anticorrelation between $\delta B_y$ and $\delta E_z$ and the
  correlation between $\delta B_z$ and $\delta E_y$ in phase are clearly
  seen in both cases, and the amplitude 
  of the magnetic field is almost identical to that of the electric
  field. It is easy to confirm that the waves carry the positive
  Poynting flux, indicating that the waves propagate toward the $+x$
  direction. All these results show that the X-mode and O-mode
  electromagnetic waves travel upstream. 
    
  \subsection{Time Evolution}\label{subsec:evo}

\begin{figure*}[htb!]
   \plottwo{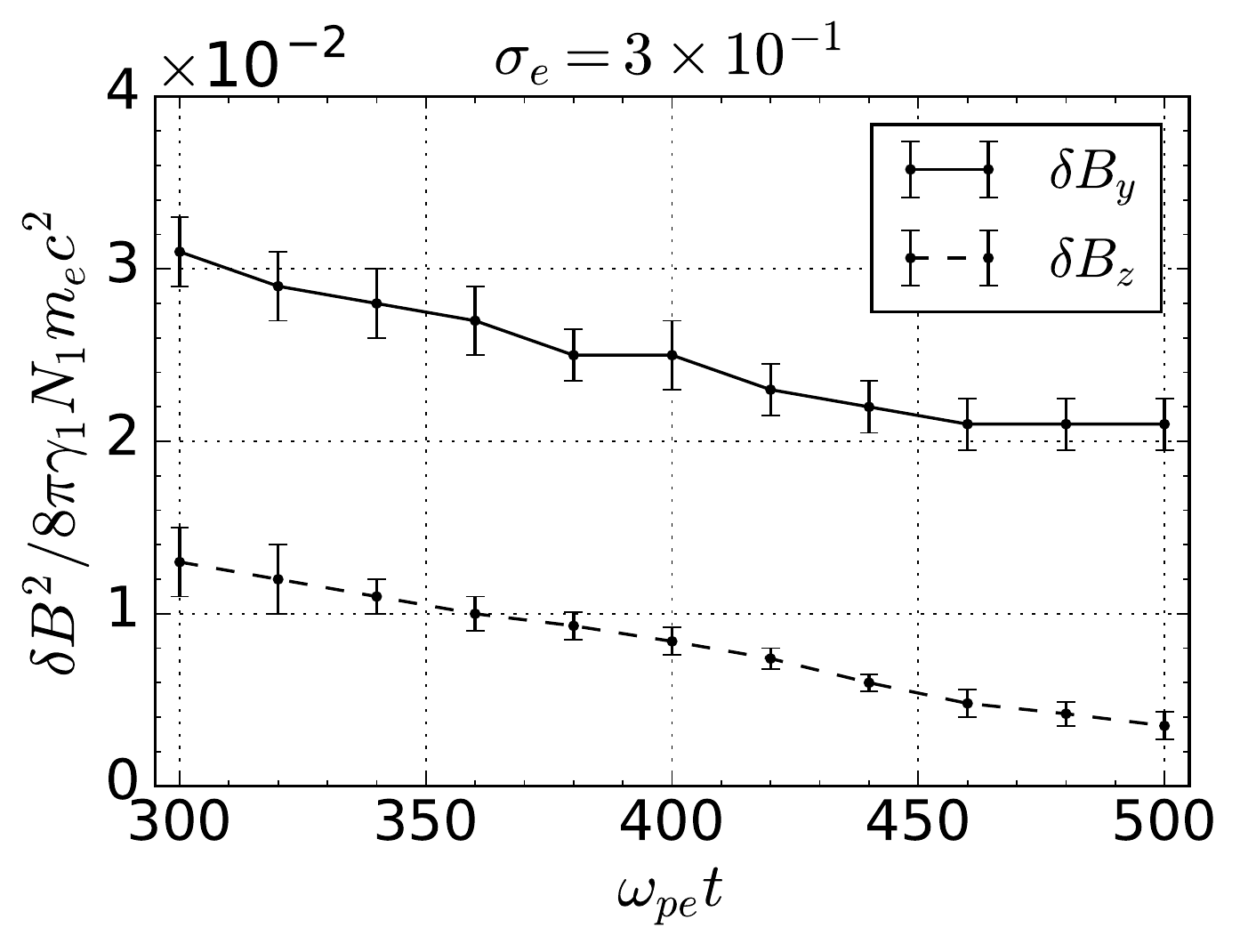}{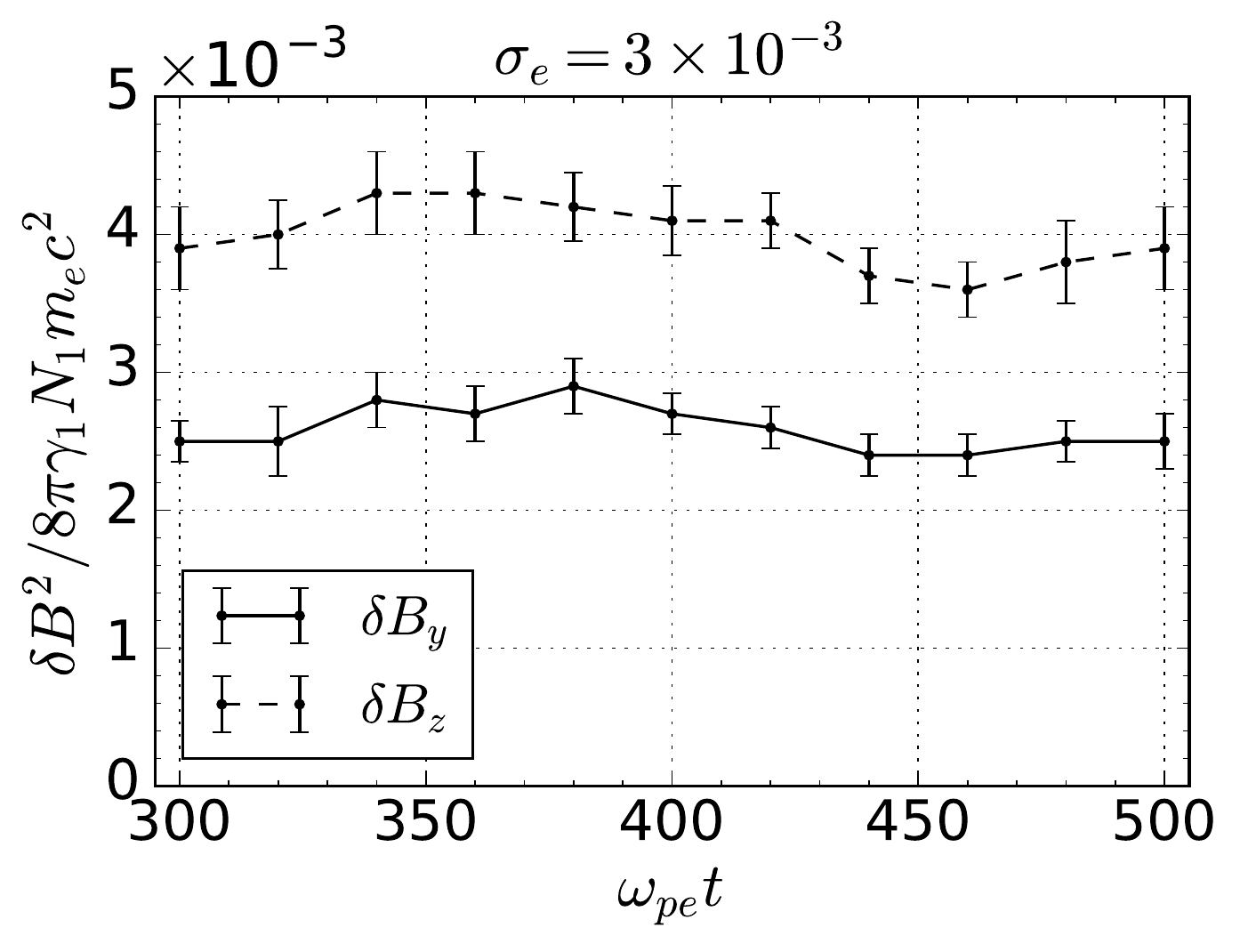}
   \caption{\replaced{Time evolution of wave energy averaged over the $y$
   direction in the precursor region given in units of upstream bulk
   kinetic energy for $\sigma_e=3\times 10^{-1}$ (left) and $\sigma_e=3\times
   10^{-3}$ (right). The solid and dashed lines indicate $y$ and $z$
   components, respectively.}{Time evolution of the average wave energy
   from $\omega_{pe}t = 300$ to $\omega_{pe}t = 500$ for
   $\sigma_e=3\times 10^{-1}$ (left) and $\sigma_e=3\times 10^{-3}$ 
   (right). The wave energy is normalized by the upstream bulk kinetic
   energy. The solid and dashed lines indicate $y$ and $z$ components,
   respectively.} }   
   \label{evo}
  \end{figure*}

    Now we discuss time evolution of the precursor wave power. Figure
  \ref{evo} shows the time evolution of the wave energy from
  $\omega_{pe}t = 300$ up to $\omega_{pe}t = 500$ for  
  $\sigma_e=3 \times 10^{-1}$ (left) and $\sigma_e=3 \times 10^{-3}$
  (right). The time evolution is determined by the same method as our
  previous study \citep{Iwamoto2017}. The wave energy is given in
  units of the upstream bulk kinetic energy, and $y$ and $z$ components
  are shown in the solid and dashed lines, respectively. 
  As shown in Figure $\ref{mode}$, the amplitude of the electric field
  is comparable to that of the magnetic field. Thus the same plots for
  the electric field is almost identical and we here show only those for
  the magnetic field.

  For $\sigma_e=3 \times 10^{-1}$, although $\delta B_y$ gradually
  declines in time, it still remains finite and gets saturated at around
  $\omega_{pe}t = 460$. In contrast, $\delta B_z$ shows continuous
  decrease. Although the O-mode wave 
  emission might be shut off after long-term evolution, the X-mode wave
  emission has already reached a quasi-steady state by the end of our
  simulation and the wave amplitude is comparable to that in the out-of-plane
  configuration (see Section \ref{subsec:dependence}). Therefore, the
  coherent electromagnetic precursor wave emission continues in the
  in-plane as well as out-of-plane configuration.

  For $\sigma_e=3\times 10^{-3}$, both $\delta B_y$ and $\delta B_z$ are
  already saturated in this time range.  Considering that $\delta B_y$
  is the component expected from the linear theory of the SMI, it is
  somewhat surprising that $\delta B_z$ is always greater than
  $\delta B_y$. We discuss the $\sigma_e$ dependence in Section
  \ref{subsec:dependence} in detail.

  \subsection{Wavenumber Spectra}\label{subsec:spectra}

\begin{figure*}
   \plottwo{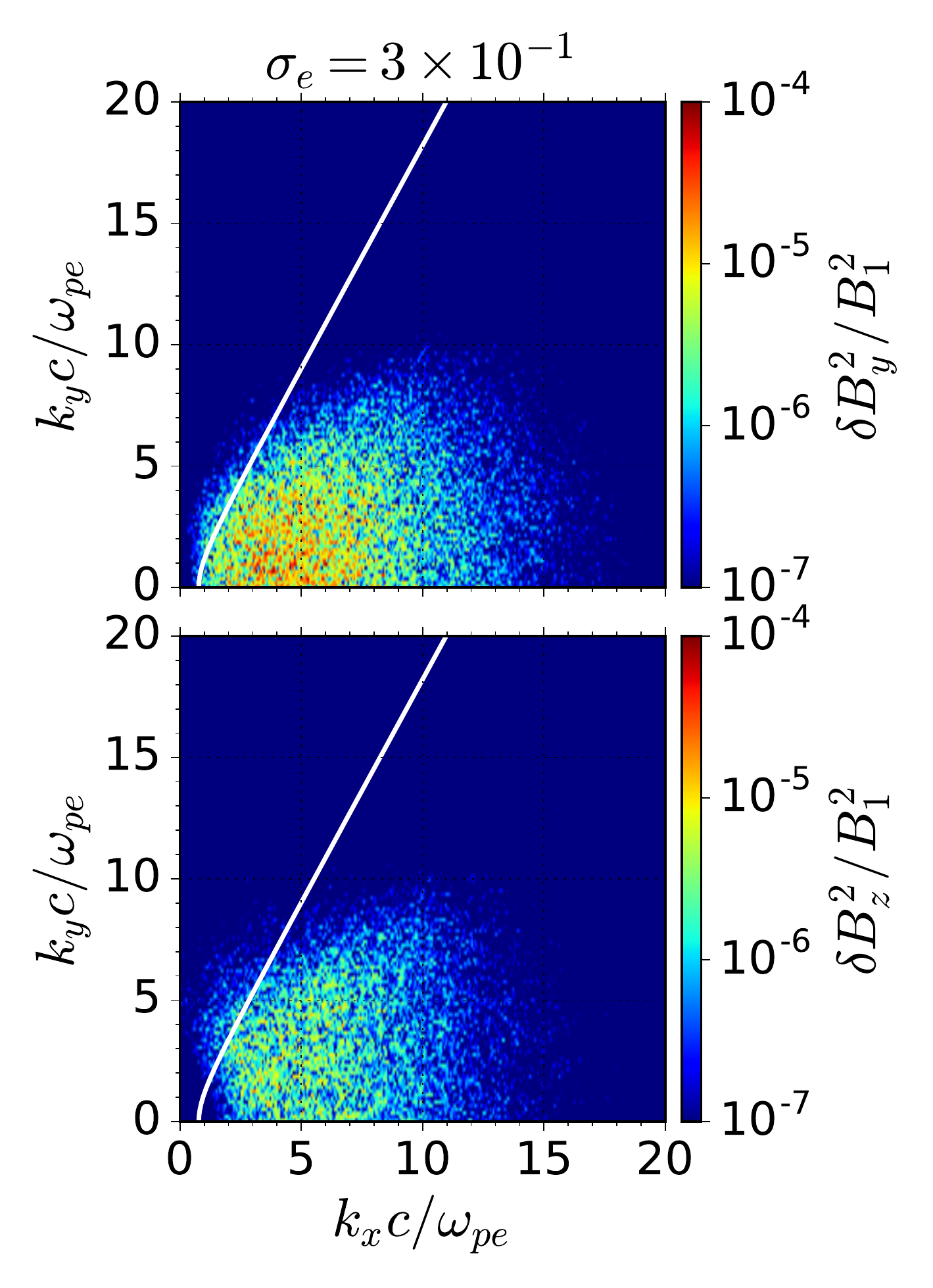}{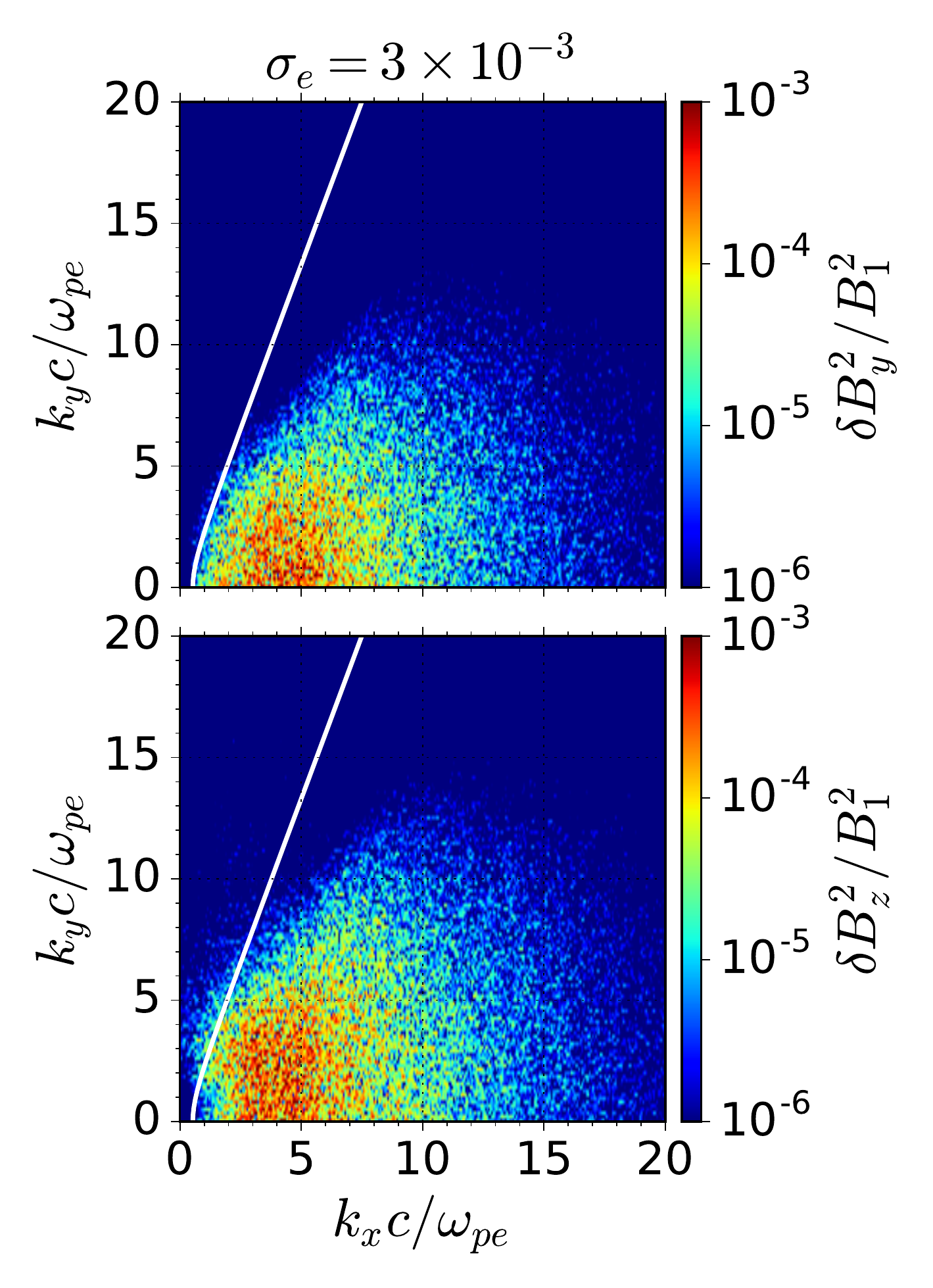}
   \caption{\replaced{Wavenumber power spectra for $y$ and $z$ components of a
   precursor wave at $\omega_{pe}t = 500$ for $\sigma_e = 3 \times
   10^{-1}$ (left column) and $\sigma_e = 3 \times 10^{-3}$ (right
   column). The white solid line indicates a theoretical cutoff
   wavenumber.}{Wavenumber power spectra of the wave
   magnetic field intensity for each component at 
   $\omega_{pe}t = 500$. The left and right column shows the spectra for 
   $\sigma_e = 3 \times 10^{-1}$ and $\sigma_e = 3 \times 10^{-3}$,
   respectively. A theoretical cutoff wavenumber is also shown in the
   white line.}} 
   \label{disp}
  \end{figure*}

  Figure \ref{disp} shows the precursor wave power spectra for each
  component in wavenumber space normalized by the upstream ambient
  magnetic field energy density. The left column shows the spectra of
  $\delta B_y$ (top) and $\delta B_z$ (bottom) for
  $\sigma_e = 3 \times 10^{-1}$, whereas the right column shows the
  spectra of  $\delta B_y$ (top) and $\delta B_z$ (bottom) for
  $\sigma_e = 3 \times 10^{-3}$. The spectra are obtained in the same
  manner as in our previous study \citep{Iwamoto2017}. Note that the
  Nyquist wavenumber for our simulation is $k_N c/\omega_{pe} \simeq
  120$ and both X-mode ($\delta B_y$) and O-mode ($\delta B_z$)
  precursor waves are well resolved.
 
  The white solid line indicates the theoretical cutoff wavenumber:
  \begin{equation}
   \label{eq:cutoff}
   k_x = \beta_{sh}\gamma_{sh}\sqrt{k_y^2+\frac{2\omega_{pe}^2}{c^2}},
  \end{equation}
  where $\beta_{sh}$ is the shock velocity normalized by the speed of
  light and $\gamma_{sh}$ is the Lorentz factor of the shock
  velocity \cite[see][Appendix B]{Iwamoto2017}. This theoretical cutoff
  wavenumber comes from the wavenumber below which the group velocity of
  the precursor wave is smaller than the shock velocity. Therefore, only
  those waves with $k_x$ greater than the threshold can escape from the
  shock toward upstream. The dispersion relation of the X-mode in a cold
  magnetized pair 
  plasma is used to derive Equation \ref{eq:cutoff}. For $\gamma_1$ and
  $\sigma_e$ used in our simulation, the dispersion relation in the
  simulation frame can be written as
  \begin{equation}
   \label{eq:omode}
   \omega^2 \simeq 2 \omega_{pe}^2 + k^2c^2.
  \end{equation}
  This dispersion relation is identical to that of the O-mode in a
  cold pair plasma, and we use Equation \ref{eq:cutoff} for the
  O-mode wave as well. The shock propagation velocity is determined from
  the time evolution of the $y$-averaged electron number density 
  $\langle N_e \rangle$, which is then used for calculation of the
  theoretical cutoff wavenumber. The result shows that the precursor
  waves are indeed propagating away from the shock, suggesting that they
  are generated at the shock front. 
  
  \subsection{$\sigma_e$ Dependence}\label{subsec:dependence}

\begin{figure*}[htb!]
   \plottwo{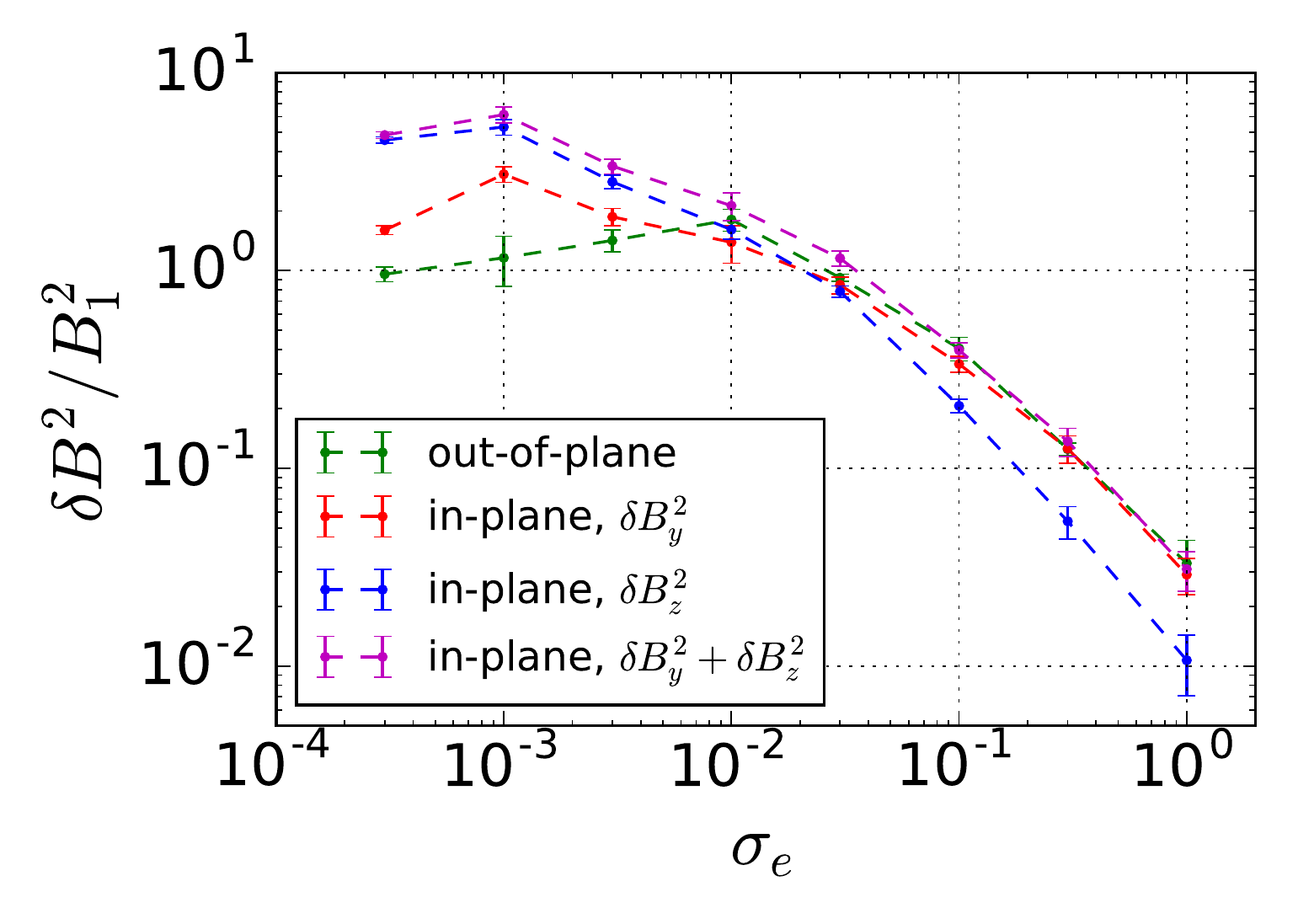}{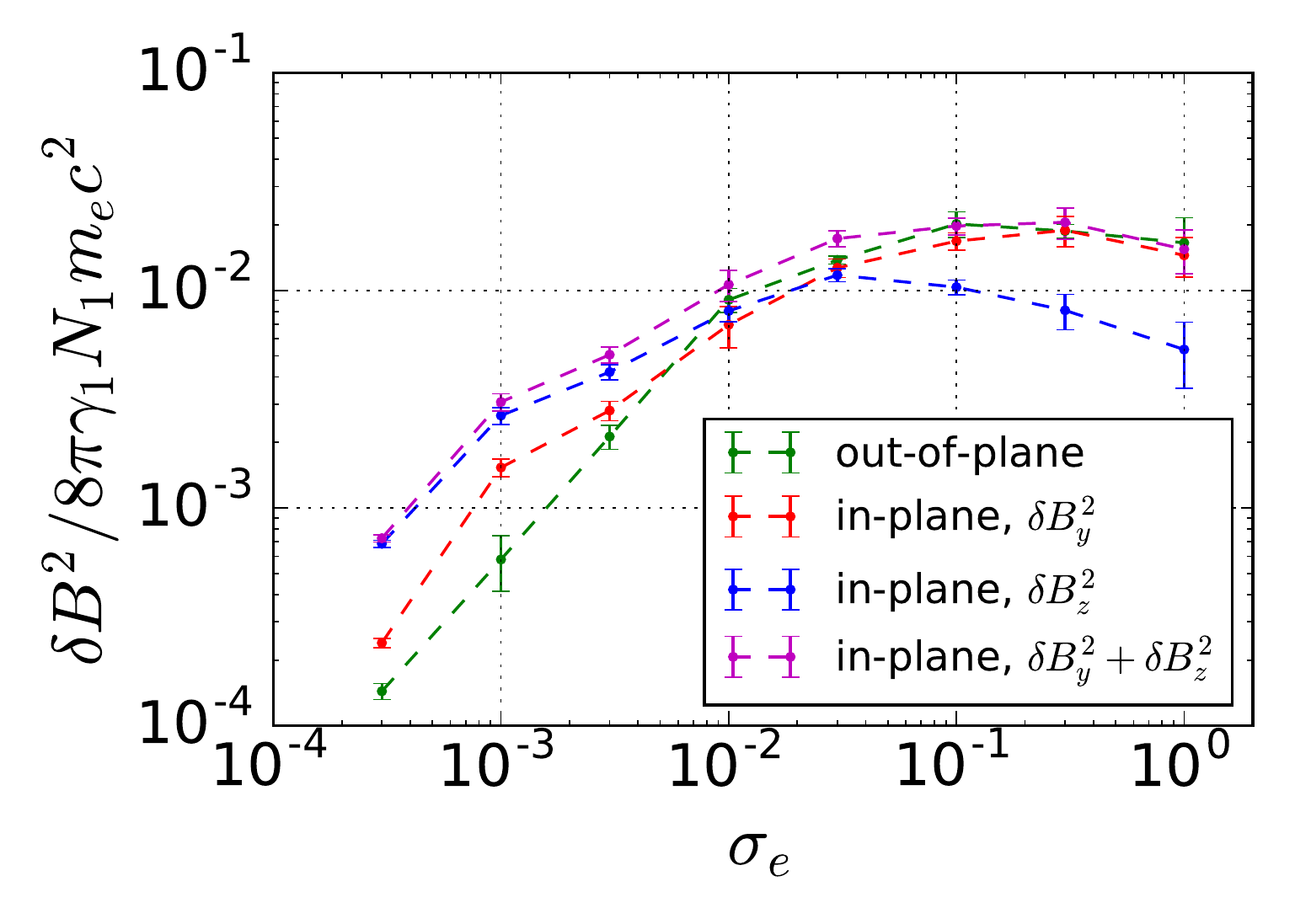}
   \caption{\replaced{$\sigma_e$ dependence of the precursor wave energy
   normalized by the ambient upstream magnetic field energy (left) and
   the upstream kinetic energy (right). }{Energy of the precursor wave
   emission given in 
   units of the ambient upstream magnetic field energy (left) and the
   upstream kinetic energy (right) as a function of $\sigma_e$.} The red,
   blue and magenta indicate $\delta B_y^2$, $\delta B_z^2$ and $\delta
   B_y^2 + \delta B_z^2$, respectively. The simulation results by
   \cite{Iwamoto2017} are shown in green.}
   \label{amp}
  \end{figure*}

  Now we discuss the $\sigma_e$ dependence of the precursor wave
  amplitude. The wave amplitude was calculated by integrating the
  power spectra (Figure \ref{disp}) over the whole wavenumber
  space. Figure \ref{amp} shows the precursor wave energy as a function 
  of $\sigma_e$ normalized by the upstream ambient magnetic field energy
  (left) and the upstream bulk kinetic energy (right). The latter may be
  understood as the energy conversion rate from the upstream bulk kinetic
  energy to the precursor wave energy. The red, blue and magenta
  indicate the X-mode wave energy $\delta B_y^2$, O-mode wave energy
  $\delta B_z^2$ and total wave energy $\delta B_y^2 + \delta B_z^2$,
  respectively. The simulation results in the out-of-plane 
  configuration by \cite{Iwamoto2017} is also shown in green for
  comparison. Note that only the X-mode precursor waves ($\delta B_z$)
  are excited in the out-of-plane configuration.

  For $\sigma_e \gtrsim 10^{-2}$, the X-mode wave energy $\delta B_y^2$
  in the in-plane configuration shows the same tendency 
  as that in the out-of-plane configuration. This may be understood as
  follows. The ambient magnetic field is larger than the magnetic field
  fluctuations generated by the instability in the shock-transition
  region and almost unperturbed for high $\sigma_e$. Thus the X-mode
  wave excitation via the SMI is nearly identical between the in-plane
  and out-of-plane configurations. 

  For $\sigma_e \lesssim 10^{-2}$, the X-mode wave energy $\delta B_y^2$
  in the in-plane configuration is greater than that in the out-of-plane
  configuration. This may be explained in terms of the coherence of the
  particle gyromotion in the shock-transition region. The WI generates
  strong magnetic field fluctuations for low $\sigma_e$ and the
  shock-transition region is dominated by the Weibel-generated magnetic
  field in both of the configurations. While charged particles, on
  average, gyrate in the $x$--$z$ plane for the in-plane
  configuration, they always gyrate in the $x$--$y$ plane for the
  out-of-plane configuration. Since the $z$ direction is ignored in our 2D
  simulations, the particle gyromotion in the in-plane case is less
  perturbed by the Weibel-generated turbulence than that in the
  out-of-plane case. In the in-plane case, therefore, the
  electromagnetic wave emission may be sufficiently amplified by the SMI
  and the wave amplitude may grow larger than that in the out-of-plane
  case.
  
  The O-mode wave energy $\delta B_z^2$ is smaller than the X-mode
  wave energy $\delta B_y^2$ for $\sigma_e \gtrsim 10^{-2}$, whereas it
  exceeds the X-mode for $\sigma_e \lesssim 10^{-2}$. This tendency
  cannot be explained by the above argument. We discuss 
  a possible excitation mechanism of O-mode waves and its relation to
  the $\sigma_e$ dependence in Section \ref{subsec:omode} in detail.  

  In conclusion, the simulation results have demonstrated that
  regardless of the orientation of the upstream ambient magnetic field,
  the precursor waves remain finite amplitude and coherent in 2D. This is
  true even for relatively low $\sigma_e$ cases where the WI grows into
  substantial amplitude in the 
  shock-transition region. The results confirm the idea that
  the coherent electromagnetic precursor wave emission is the real
  nature of the relativistic magnetized shocks.

 \section{Discussion}\label{sec:discussion}

  \subsection{Excitation Mechanism of O-mode Waves}\label{subsec:omode}

  Our simulation results show that the O-mode as well as X-mode
  electromagnetic waves are excited in relativistic shocks. As we
  already mentioned, the linear 
  theory of the SMI \citep{Wu1979, Lee1980, Melrose1982, Melrose1984}
  predicts that the X-mode wave emission overwhelms the
  O-mode wave emission. However, in the in-plane configuration, the O-mode
  precursor wave is clearly identified in the upstream region. This may be
  explained qualitatively as follows.   

  In the early stage of the simulation, the ambient magnetic field in
  the shock transition is entirely oriented along the $y$ axis. Charged
  particles gyrate in the $x$--$z$ plane and induce the SMI. Since the
  X-mode wave has a fluctuating component of the magnetic field parallel
  to the ambient magnetic field, only $\delta B_y$ is excited in this
  stage. When the fluctuations in $B_z$ generated by the plasma
  instabilities in the shock-transition region have grown to be a
  non-negligible fraction of the ambient magnetic field, the net ambient
  magnetic field in the shock-transition region is undulated in the
  $y$--$z$ plane. If the SMI is induced by particles gyrating around 
  the net ambient magnetic field, the X-mode wave in the shock-transition
  region will have $\delta B_z$ as well as $\delta B_y$. Such a X-mode
  wave experiences changes in the direction of the ambient magnetic
  field during its propagation toward upstream. If the polarization of
  the wave electromagnetic field remains unchanged during the
  propagation,  the X-mode wave may be mode-converted into an O-mode
  wave in the upstream. By performing simple PIC simulations, we have
  confirmed that this hypothesis is indeed correct. That is, a X-mode
  wave keeps its polarization and is converted into an O-mode as it
  propagates through a layer of magnetic field rotation. Therefore, we
  believe that O-mode waves observed in the precursor region are the
  result of mode conversion from the X-mode generated by the SMI in the
  turbulent shock-transition region. 

  The delay of the O-mode wave in our simulation provides indirect
  evidence for this model. The excitation mechanism shows that the
  O-mode wave is excited after the generation of the strong magnetic
  field fluctuations by the instabilities in the shock-transition
  region. In fact, the generation time of the O-mode waves
  ($\omega_{pe}t \sim 80$ for $\sigma_e = 3 \times 10^{-1}$ and
  $\omega_{pe}t \sim 40$ for $\sigma_e = 3 \times 10^{-3}$) is roughly
  identical to the saturation time of the plasma instabilities (see
  Appendix \ref{sec:linear}).

  The $\sigma_e$ dependence of the O-mode wave amplitude in Figure
  \ref{amp} may be explained by this excitation mechanism. For
  $\sigma_e \gtrsim 10^{-2}$, the ambient magnetic field is much larger
  than the magnetic field fluctuations and almost aligned in the $y$
  direction. Thus $\delta B_y$ should be the main component of the X-mode
  wave in the shock-transition region, and $\delta B_z$ which is observed as
  the O-mode wave in the upstream region may be much smaller for high
  $\sigma_e $.

  For $\sigma_e \lesssim 10^{-2}$, the Weibel-generated magnetic field
  dominates over the ambient magnetic field and the effective $\sigma_e$
  becomes much larger in the shock-transition region. The higher
  effective $\sigma_e$ allows a wave generated via a lower-order
  cyclotron harmonic resonance $n$ to satisfy the condition 
  $\omega = n \omega_{ce} \gtrsim
  \sqrt{2(1+\beta_{sh}\gamma_{sh})}\omega_{pe}$
  such that it can propagate upstream \citep[see][]{Iwamoto2017}. We
  think that the lower cyclotron harmonics contribution may be the
  reason for the enhanced power of $\delta B_z$. In other words, the
  Weibel-generated magnetic field plays the role for the enhanced O-mode
  wave power. This model indicates that the O-mode wave continues to
  exist with a finite amplitude for considerably lower $\sigma_e$.

  \subsection{Implication for 3D}\label{subsec:3d}

  Based on the 2D simulation results obtained with both the in-plane and
  out-of-plane configurations, we now discuss implication for
  three-dimensional (3D) systems. As
  discussed in Section \ref{subsec:omode}, the O-mode precursor wave 
  emission is attributed to the large-amplitude magnetic field
  fluctuations generated in the shock-transition region, in particular
  by the WI at low $\sigma_e$. Since we can naturally expect the
  presence of such fluctuations in 3D, the O-mode precursor waves 
  will also be excited. It is, however, not easy to estimate the
  relative emission efficiency between the O-mode and X-mode. Since the
  particle gyromotion in 3D should be less coherent than the 2D with the
  in-plane configuration, the O-mode wave power will become
  smaller. Concerning the gyromotion in the shock-transition region, the
  out-of-plane configuration may better represent 3D.
  
  We have found that the particle acceleration efficiency also depends
  on the magnetic field configuration. Non-thermal particles are
  not generated in the in-plane configuration in 2D (see Appendix
  \ref{sec:energy}), whereas a clear non-thermal tail is observed in the
  energy spectra for low $\sigma_e$ in the out-of-plane configuration
  \citep[see][Figure 10]{Iwamoto2017}. Considering the suppression of
  the cross-field diffusion in the in-plane configuration, again the 
  out-of-plane configuration may be closer to 3D concerning the particle
  acceleration efficiency.
  
  In any case, the important fact is that the intense coherent precursor
  wave can be excited for a wide range of $\sigma_e$ in both of the
  configurations. This strongly indicates that the intense coherent 
  precursor wave emission is intrinsic to relativistic magnetized shocks
  and even in 3D.

  \subsection{Implication for WFA in Relativistic
  Shocks}\label{subsec:wfa}
  
\begin{figure}[htb!]
   \plotone{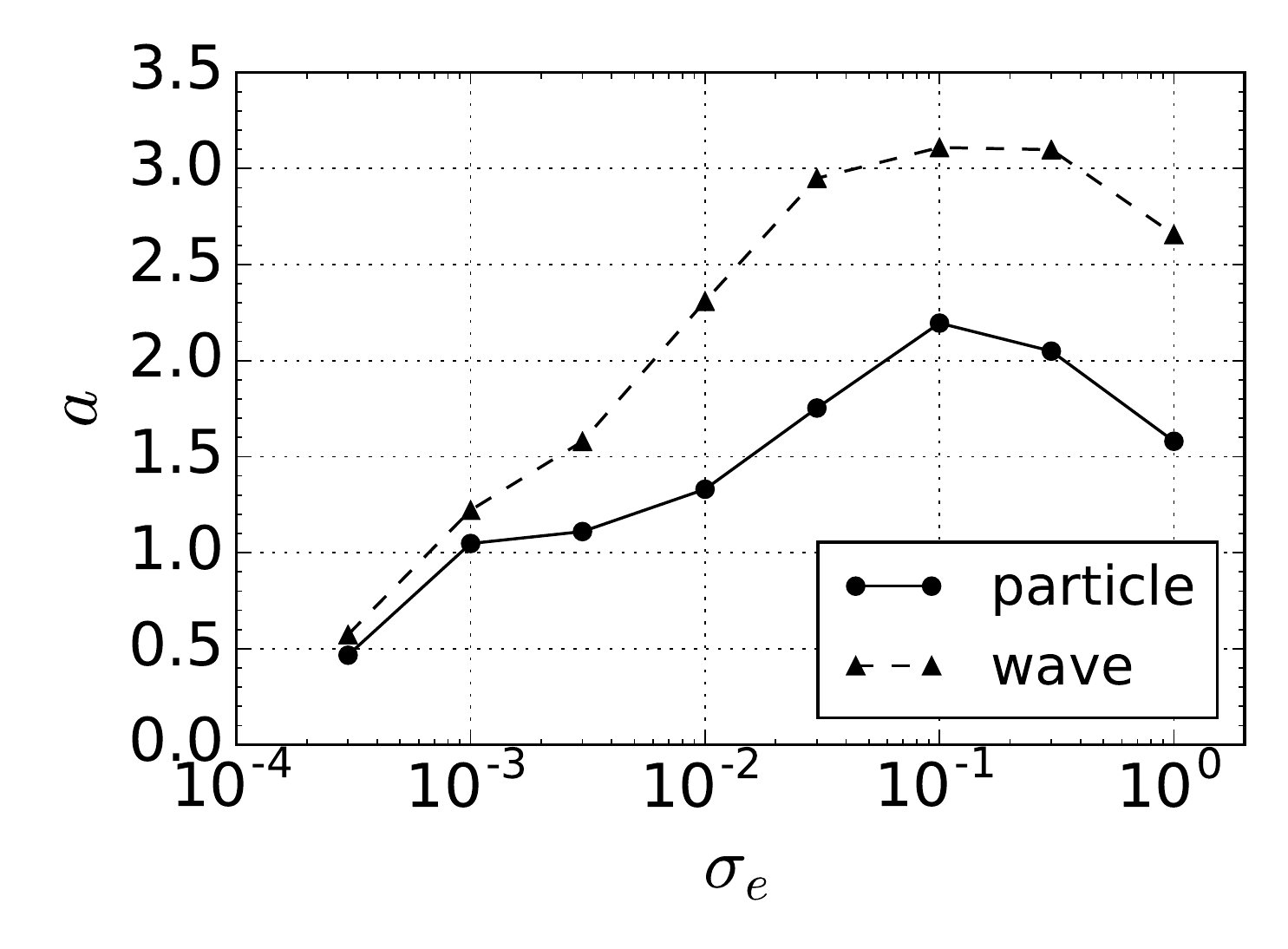}
   \caption{\replaced{$\sigma_e$ dependence of the strength parameters
   estimated from the amplitudes of electron quiver motion (solid line)
   and the precursor wave amplitude (dashed line)}{Strength parameter of
   the precursor wave as a function of $\sigma_e$. The solid and dashed
   line indicate the estimation method based on the particle quiver
   velocity and the wave amplitude, respectively.}}
   \label{strength}
  \end{figure}

  Now we discuss the feasibility of the WFA in relativistic
  shocks. The WFA requires an intense electromagnetic wave in the sense
  that the wave strength parameter $a = e\delta E/m_ec\omega$ is greater
  than unity, where 
  $\delta E$ is the amplitude of the wave electric field and $\omega$ is
  the wave frequency \citep{Kuramitsu2008}. We estimated the strength
  parameter of the precursor wave with two different methods; one based
  on the oscillation amplitude of the transverse particle velocity, the
  other based on the wave amplitude. The details can be found in our
  previous paper \citep{Iwamoto2017} except that the total wave power
  $\sqrt{\delta B_y^2 + \delta B_z^2}$ is used here. The results are
  shown in Figure \ref{strength} which demonstrates that the amplitudes
  of the precursor waves are indeed quite large.  
  
\begin{figure}[htb!]
   \plotone{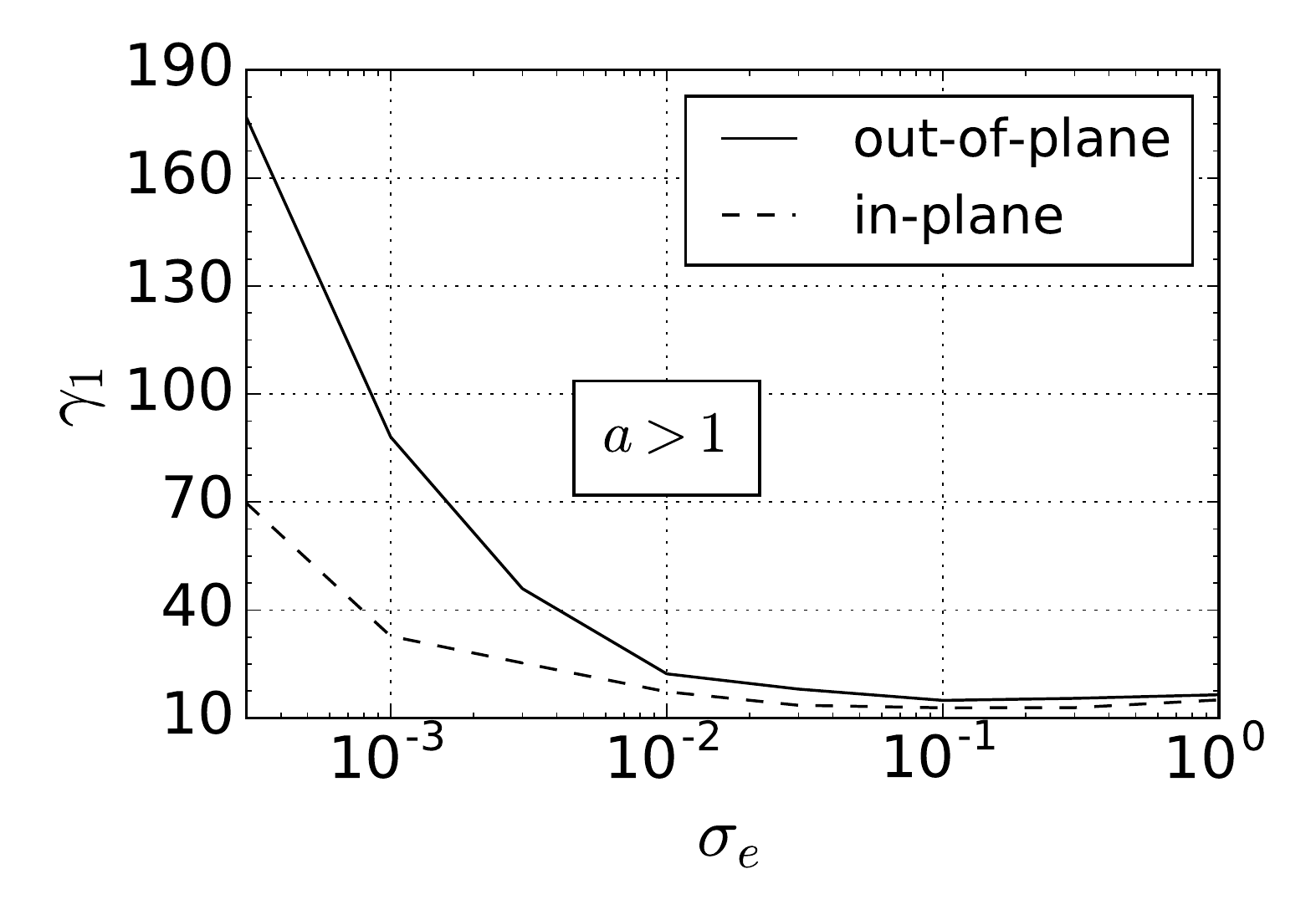}
   \caption{Parameter space plot in $\sigma_e$ and $\gamma_1$. The solid
   and dashed line indicates the out-of-plane and in-plane case,
   respectively. Strength parameter $a$ is greater 
   than unity in the region above each line. } 
   \label{sig-gam}
  \end{figure}

  Assuming a linear scaling of the strength parameter with respect to
  the Lorentz factor $\gamma_1$ \citep[see, e.g.,][]{Hoshino2008}, we may
  estimate the region in the 
  $\sigma_e$--$\gamma_1$ parameter space where the WFA is effective. In
  Figure \ref{sig-gam}, the solid and dashed line indicates the estimates
  obtained by using the simulation results for the out-of-plane and
  in-plane configuration, respectively. This clearly indicates that
  higher Lorentz factors and moderate magnetizations are favorable for
  the WFA model. Again, we draw the same conclusion as
  \cite{Iwamoto2017} that highly relativistic external shocks of GRBs
  are candidate sites for acceleration of ultra-high-energy cosmic rays. 

  The above discussion primarily focused on the wave amplitude. Although
  the large-amplitude precursor waves will induce wakefield in
  ion--electron plasmas, it is not clear yet whether the wakefield can  
  sufficiently accelerate particles. The previous study demonstrated
  the WFA using Gaussian laser pulse \citep{Kuramitsu2008}. However, the
  actual precursor waves are a superposition of waves continuously
  emitted from different positions in the shock front, and the spectra
  are rather broadband in wavenumber as shown in Figure
  \ref{disp}. A ponderomotive force exerted by such waves should become
  weaker \citep[e.g.,][]{Kruer1988,Hoshino2008}. Therefore, the
  generation of wakefield and particle acceleration  may be less
  efficient. \added{Also, different properties of the WI in pair and
  ion-electron plasmas, energy exchange between ions and electrons
  \citep[e.g.,][]{Kumar2015} may influence the efficiency of the
  particle acceleration.} These issues should be examined by performing
  shock simulations in ion--electron plasmas in the future.

  \section{Summary}\label{sec:summary}
  
  In this work, we performed 2D simulations of relativistic
  perpendicular shocks in pair plasmas with the in-plane ambient
  magnetic field, and investigated the
  physics of the intense coherent precursor wave emission. In the
  in-plane configuration, O-mode as well as X-mode electromagnetic
  precursor waves are excited. We think that the O-mode waves are
  initially excited as X-mode by the SMI in the shock-transition
  region. Since the instabilities in the shock-transition region
  generate fluctuations in $B_z$ and disturb the ambient magnetic field,
  the SMI should excite X-mode waves which have $\delta B_z$ as well as
  $\delta B_y$. The generated waves having $\delta B_z$ may be
  mode-converted into O-mode during the propagation to the upstream
  region. The delay of the O-mode wave identified in our simulation is
  consistent with this model. We quantified the precursor wave amplitude
  as a function of the magnetization parameter $\sigma_e$ and compared
  the simulation results with that in the out-of-plane configuration by
  \cite{Iwamoto2017}. The wave amplitude is sufficiently large to
  disturb the upstream plasma even in the Weibel-dominated regime and the
  transverse density filaments are generated as in the case of the
  out-of-plane configuration. We thus conclude that the precursor wave
  emission is the real nature of the realistic magnetized shocks.
  
  In the range of $\sigma_e$ used in our
  simulations, the precursor wave keeps coherent and its amplitude is
  large enough to induce the wakefield. Therefore, the WFA may operate
  in relativistic ion--electron shocks. 

  \acknowledgments
  
  We are grateful to Jacek Niemiec, Martin Pohl, Oleh Kobzar, Arianna
  Ligorini and  Artem Bohdan for fruitful discussion.

  Numerical computations and analyses were in part carried out on Cray
  XC30 and computers at Center for Computational Astrophysics, National
  Astronomical Observatory of Japan.

  This work used the computational resources of the K computer provided
  by the RIKEN Advanced Institute for Computational Science through the
  HPCI System Research Project (Project ID: hp150263), and was supported
  in part by JSPS KAKENHI Grant Number 17H02877 
  
  This work also used the computational resources of the HPCI system
  provided by Information Technology Center, Nagoya University through the
  HPCI System Research Project (Project ID: hp170158).
  
  \appendix

 \section{Electromagnetic Instabilities in Shock-Transition
 Region}\label{sec:linear} 

 Here we present linear analysis of the electromagnetic instabilities
 excited in the shock-transition region. The dispersion relation for
 electromagnetic waves propagating parallel to an ambient magnetic field
 is given by 
 \citep[see, e.g.,][]{Yoon1987}
 \begin{equation}
  \label{eq:weibel1}
   D(k,\omega) = 1-\frac{c^2k^2}{\omega^2}
   +\sum_s\frac{\Omega_{ps}^2}{\omega^2}\int_{-\infty}^{\infty}\int_{0}^{\infty}
   \frac{(\gamma\omega - cku_{\parallel})\partial F_{0s}/\partial
   u_{\perp} +cku_{\perp}\partial F_{0s}/\partial
   u_{\parallel}}{\gamma\omega \pm\Omega_{cs}-cku_{\parallel}}\frac{\pi
   u_{\perp}^2}{\gamma}du_{\perp}du_{\parallel},
 \end{equation}
 where $\gamma = \sqrt{1+u_{\perp}^2+u_{\parallel}^2}$ is the Lorentz
 factor, $\Omega_{ps}$ is the non-relativistic plasma
 frequency, $\Omega_{cs}$ is the non-relativistic
 cyclotron frequency and $F_{0s}$ is the unperturbed distribution
 function normalized as follows:  
 \begin{equation}
  \label{eq:weibel2}
  \int_{-\infty}^{\infty}\int_{0}^{\infty}
   F_{0s}(u_{\parallel},u_{\perp})2\pi u_{\perp}du_{\perp}du_{\parallel} = 1.
 \end{equation}
 The subscript $s$ indicates particle species (i.e., electron and positron).
 In Equation \ref{eq:weibel1}, the positive (negative) sign corresponds to
 the right-hand (left-hand) polarization. 
  
 We assume a cold ring distribution for both electrons and positrons,
 \begin{equation}
   \label{eq:weibel3}
  F_{0s} = \frac{1}{2\pi u_{0s}}\delta(u_{\perp}-u_{0s})\delta(u_{\parallel}).
 \end{equation}
 By substituting Equation \ref{eq:weibel3} for Equation
 \ref{eq:weibel1}, we obtain 
 \begin{equation}
   \label{eq:weibel4}
   D(k,\omega) =
   1-\frac{c^2k^2}{\omega^2}
   -\sum_s\frac{\omega_{ps}^2}{\omega(\omega\pm\omega_{cs})}
   +\frac{1}{2}\left(1-\frac{c^2k^2}{\omega^2}\right)
   \sum_s\left(1-\frac{1}{\gamma_{0s}^2}\right)\frac{\omega_{ps}^2}
   {(\omega\pm\omega_{cs})^2},
 \end{equation}
 where $\gamma_{0s}=\sqrt{1+u_{0s}^2}$ is the initial Lorentz factor,
 $\omega_{ps}=\Omega_{ps}/\sqrt{\gamma_{0s}}$ is the relativistic plasma
 frequency and $\omega_{cs}=\Omega_{cs}/\gamma_{0s}$ is the relativistic
 cyclotron frequency. When $\gamma_{0s} = 1$,
 Equation \ref{eq:weibel4} is identical to the dispersion relation in a
 cold magnetized plasma. Introducing $\omega_{pe} \equiv
 \omega_{pe^-}=\omega_{pe^+}$ and $\omega_{ce} \equiv -\omega_{ce^-} =
 \omega_{ce^+} > 0$, Equation \ref{eq:weibel4} reduces
 \begin{equation}
 \label{eq:weibel5}
  D(k,\omega) =
   1-\frac{c^2k^2}{\omega^2}-
   \frac{2\omega_{pe}^2}{\omega^2-\omega_{ce}^2}
   +\beta_0^2\left(1-\frac{c^2k^2}{\omega^2}\right)
   \frac{\omega_{pe}^2(\omega^2+\omega_{ce}^2)}{(\omega^2-\omega_{ce}^2)^2},
 \end{equation}
 where $\beta_0 = u_{0e^{\pm}}/\gamma_{0e^{\pm}}$.
 $D(k,\omega)=0$ is expressed as follows:
 \begin{equation}
  \label{eq:weibel6}
   \left(\frac{\omega}{\omega_{pe}}\right)^6-
   \left(\frac{c^2k^2}{\omega_{pe}^2}+2\sigma_e-\beta_0^2+2\right)
   \left(\frac{\omega}{\omega_{pe}}\right)^4 \\ 
   +\left[(2\sigma_e-\beta_0^2)\frac{c^2k^2}{\omega_{pe}^2}
   +\sigma_e(\sigma_e+\beta_0^2+2)\right]
   \left(\frac{\omega}{\omega_{pe}}\right)^2-
   \sigma_e(\sigma_e+\beta_0^2)\frac{c^2k^2}{\omega_{pe}^2}=0,
 \end{equation}
 where $\sigma_e = \omega_{ce}^2/\omega_{pe}^2$.
 In this linear analysis, we assume that the wavenumber $k$ is a real
 number and consider only 
 the region where $k \ge 0$ and $\omega \ge 0$ 
 because of the symmetry of $k$ and $\omega$ for a pair plasma.  

 First, we study the case of $\sigma_e > (2-\beta_0^2)^2/8\beta_0^2$.
 $\omega$ becomes a complex number at $k \le k_1$ and $k \ge k_2$.
 Here, the threshold wavenumber $k_{1}$ and $k_2$ are determined from
 \begin{eqnarray}
  \left(\frac{ck_1}{\omega_{pe}}\right)^2 =
   f\left(\frac{\omega_1^2}{\omega_{pe}^2}\right), \\
  \left(\frac{ck_2}{\omega_{pe}}\right)^2 =
   f\left(\frac{\omega_2^2}{\omega_{pe}^2}\right), \\
  k_1 < k_2, \\
  f(x) = x - 2 - \frac{2(\sigma_e-\beta_0^2)x-2\sigma_e(\sigma_e+\beta_0^2)}
   {x^2-(2\sigma_e-\beta_0^2)x+\sigma_e(\sigma_e+\beta_0^2)}.
 \end{eqnarray}

\begin{figure}[htb!]
  \plotone{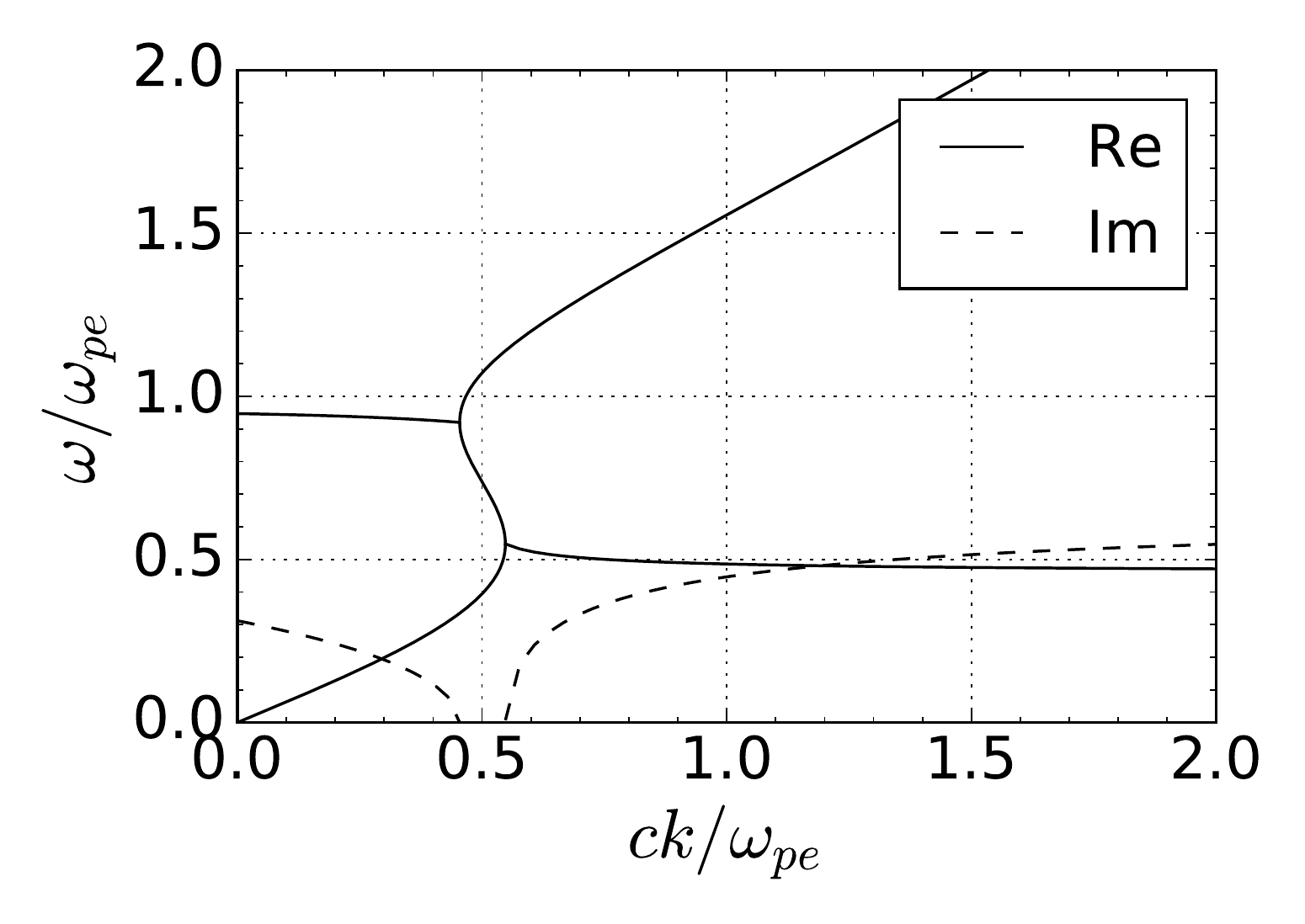}
  \caption{Dispersion relation for $\sigma_e = 0.3$ and
  $\gamma_0=40$. The solid and dashed line indicates the real and
  imaginary part of the frequency, respectively.}  
  \label{wk_lowsig}
 \end{figure}

 $\omega_1$ and $\omega_2$ satisfy
 \begin{eqnarray}
  f^{\prime}\left(\frac{\omega_1^2}{\omega_{pe}^2}\right)=
   f^{\prime}\left(\frac{\omega_2^2}{\omega_{pe}^2}\right) = 0, \\
  \omega_1 > \omega_2.
 \end{eqnarray}
 Figure \ref{wk_lowsig} shows the dispersion relation with $\sigma_e =
 0.3, \gamma_0=40$ numerically obtained from Equation
 \ref{eq:weibel6}. The real and imaginary part of 
 the frequency is shown by the solid and dashed line, respectively. The
 unstable branch for $k \le k_1$ and $k \ge k_2$ is connected to
 the electromagnetic and Alfv\'en mode branch, respectively. We think
 the unstable mode for $k \ge k_2$ corresponds to the
 Alfv\'en-ion-cyclotron instability in ion--electron plasmas. The 
 growth rate of the mode $k \le k_1$ has its maximum at $k = 0$, 
 \begin{eqnarray}
  \frac{{\rm Re}(\omega_{max})}{\omega_{pe}}  &=&
   \sqrt{\frac{1}{2}\sigma_e-\frac{1}{4}\beta_0^2+\frac{1}{2}
   +\frac{1}{2}\sqrt{\sigma_e(\sigma_e+\beta_0^2+2)}}, \\
  \frac{{\rm Im}(\omega_{max})}{\omega_{pe}} &=&
  \sqrt{-\frac{1}{2}\sigma_e+\frac{1}{4}\beta_0^2-\frac{1}{2}
  +\frac{1}{2}\sqrt{\sigma_e(\sigma_e+\beta_0^2+2)}}.
 \end{eqnarray}
 The maximum growth rate of the mode $k \ge k_2$ occurs for
 $ck/\omega_{pe} \gg 1$,
 \begin{eqnarray}
  \label{eq:real1}
  \frac{{\rm Re}(\omega_{max})}{\omega_{pe}}  &=&
   \sqrt{\frac{1}{2}\sigma_e-\frac{1}{4}\beta_0^2
   +\frac{1}{2}\sqrt{\sigma_e(\sigma_e+\beta_0^2)}}, \\
  \label{eq:growth1}
   \frac{{\rm Im}(\omega_{max})}{\omega_{pe}} &=&
  \sqrt{-\frac{1}{2}\sigma_e+\frac{1}{4}\beta_0^2
  +\frac{1}{2}\sqrt{\sigma_e(\sigma_e+\beta_0^2)}}.
 \end{eqnarray}
 If $\sigma_e \gg 1$, then the maximum growth rates for both modes
 are written as follows:
 \begin{eqnarray}
  {\rm Re}(\omega_{max}) &\sim& \omega_{ce}, \\
  {\rm Im}(\omega_{max}) &\sim& \frac{\beta_0}{\sqrt{2}}\omega_{pe}.
 \end{eqnarray}
 
\begin{figure}[htb!]
  \plotone{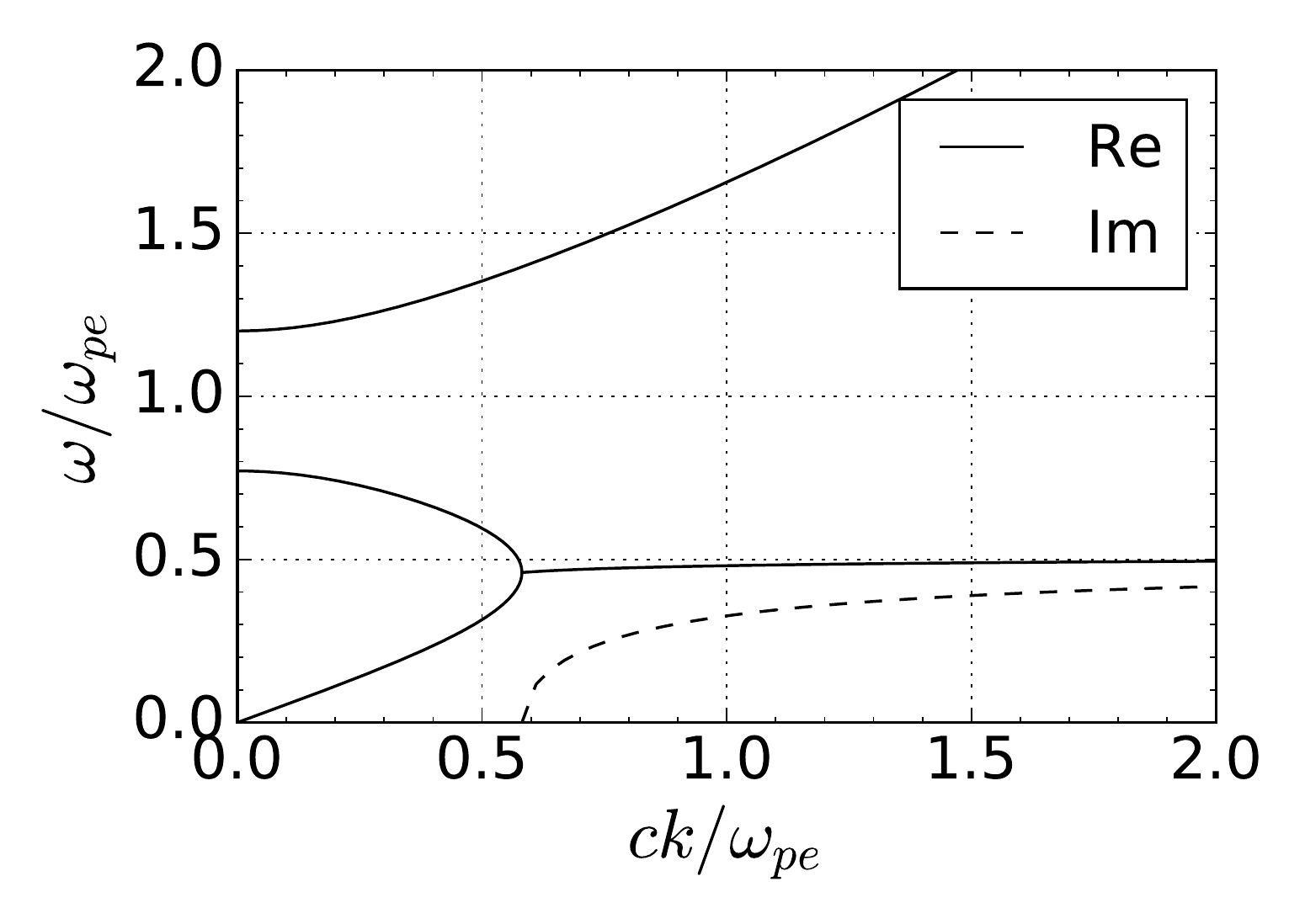}
  \caption{Dispersion relation for $\sigma_e = 0.3$ and
  $\gamma_0=2$.}  
  \label{wk_midsig}
 \end{figure}

 Second, we study the case of 
 $\beta_0^2/8 \le \sigma_e \le (2-\beta_0^2)^2/8\beta_0^2$.
 For highly relativistic plasma $\beta_0 \sim 1$, the condition is
 satisfied within a very narrow range of $\sigma_e$. In this case, only the
 instability on the Alfv\'en mode branch exists at $k \ge k_2$. The
 dispersion relation is shown in Figure \ref{wk_midsig}. The maximum
 growth rate is identical to Equation \ref{eq:real1} and \ref{eq:growth1}.
 
\begin{figure}[htb!]
  \plotone{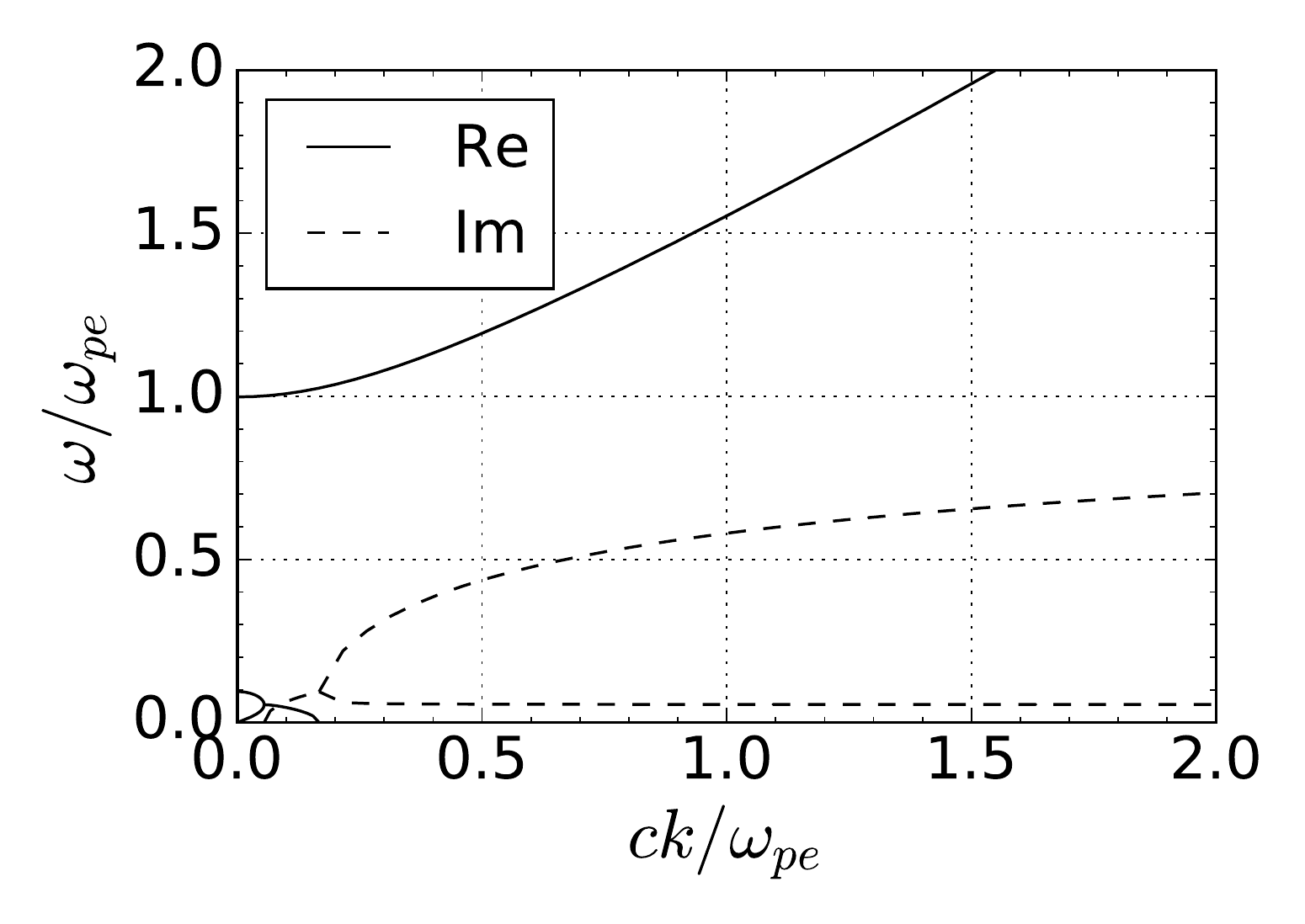}
  \caption{Dispersion relation for $\sigma_e=3\times10^{-3}$ and
  $\gamma_0=40$.}   
  \label{wk_highsig} 
 \end{figure}

 Finally, we study the case of $\sigma_e < \beta_0^2/8$. $\omega$
 is a complex number and a pure imaginary number for $k_2 \le k < k_3$
 and $k \ge k_3$, respectively. Here, the  
 threshold wavenumber $k_3$ is determined by 
 \begin{equation}
  \left(\frac{ck_3}{\omega_{pe}}\right)^2
   =f\left(\frac{\omega_3^2}{\omega_{pe}^2}\right).
 \end{equation}
 The threshold frequency $\omega_3$ satisfies
 \begin{eqnarray}
  f^{\prime}\left(\frac{\omega_3^2}{\omega_{pe}^2}\right)=0, \\
  \omega_-<{\rm Im}(\omega_3)<\omega_+, \\
  {\rm Re}(\omega_3)=0,
 \end{eqnarray}
 where
 \begin{equation}
  \frac{\omega_{\pm}}{\omega_{pe}}
   =\sqrt{-\sigma_e+\frac{1}{2}\beta_0^2 \pm
   \frac{1}{2}\beta_0\sqrt{\beta_0^2-8\sigma_e}}.
 \end{equation}
 Figure \ref{wk_highsig} shows the dispersion relation with $\sigma_e =
 3\times10^{-3}, \gamma_0=40$. 
 For $\sigma_e \le \beta_0^2/8$, there are two purely growing modes when
 the wavenumber is greater than the threshold wave 
 number $k_3$. We think the upper unstable branch corresponds to the
 WI. The growth rates of these modes asymptotically approach 
 $\omega_{\pm}$ as the wavenumber increases. The maximum growth rate is
 expressed as follows:
 \begin{equation}
  \label{eq:growth2}
  {\rm Im}(\omega_{max}) \sim \omega_+.
 \end{equation}
 If $\sigma_e \ll \beta_0^2$, then the maximum growth rate is
 written as follows:
 \begin{equation}
 {\rm Im}(\omega_{max}) \sim \beta_0\omega_{pe}.
 \end{equation}

 \begin{figure*}[htb!]
  \plottwo{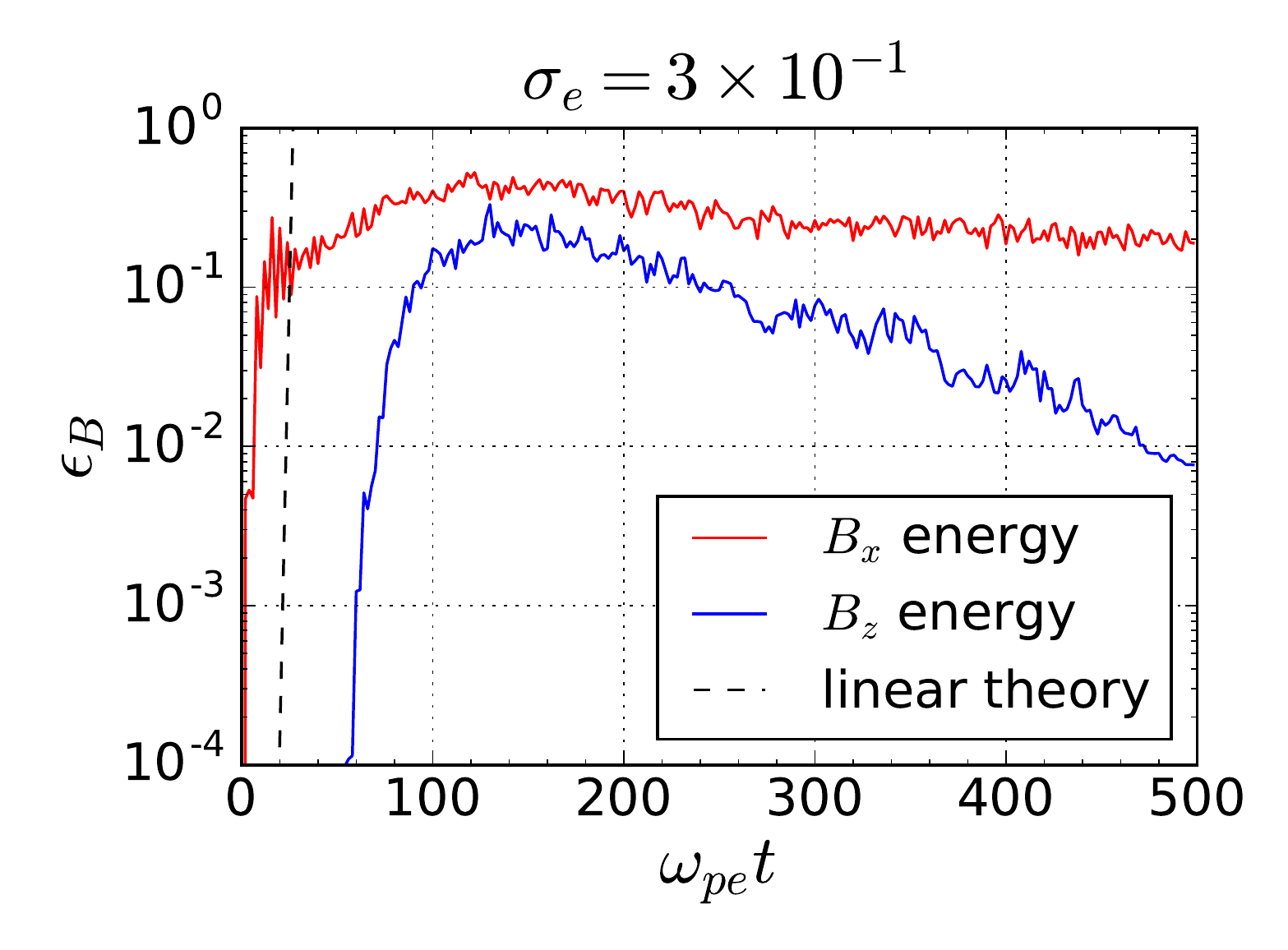}{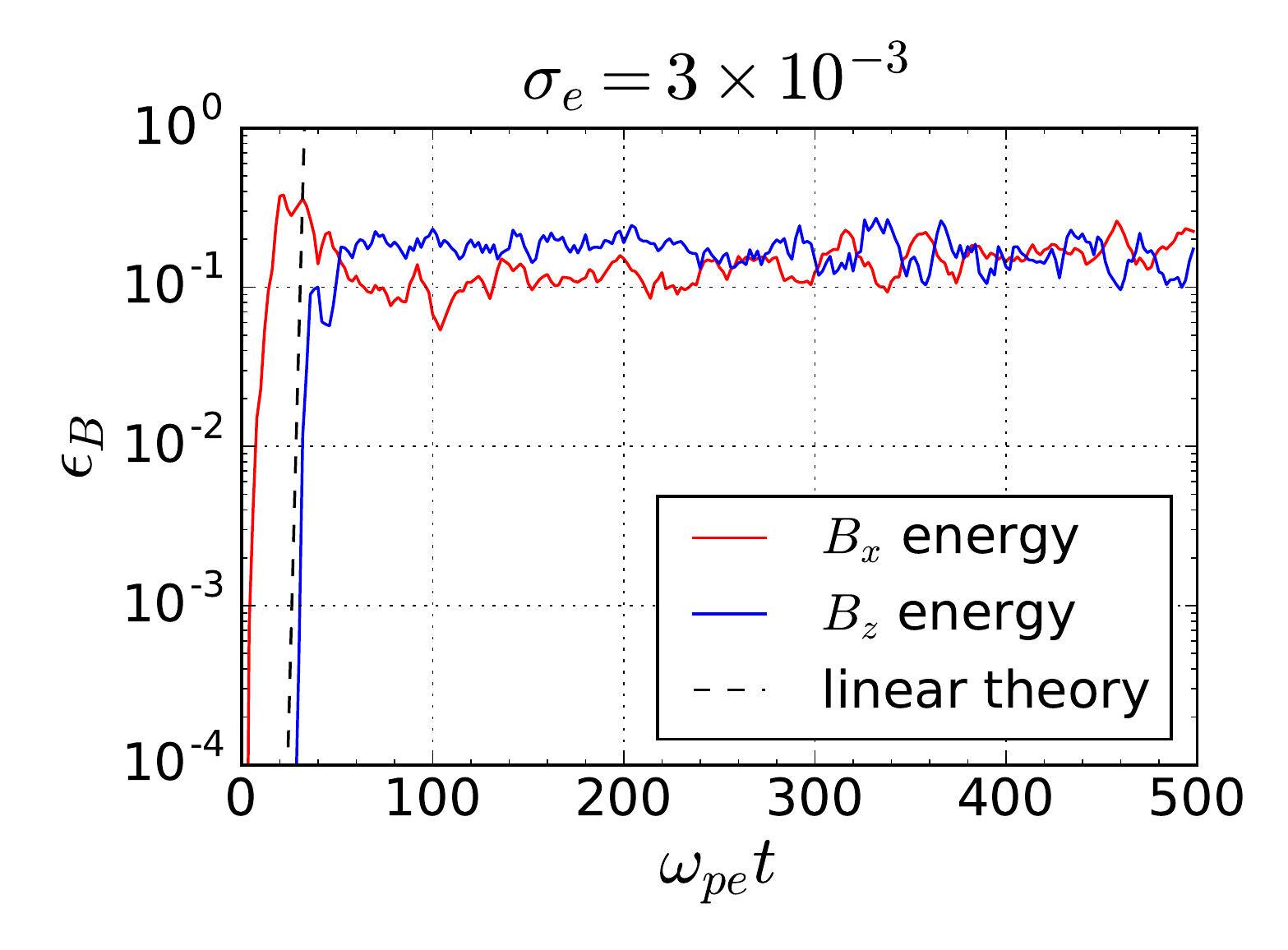}
  \caption{Time evolution of the fluctuating magnetic field energy
  normalized by the upstream bulk kinetic energy for $\sigma_e = 3 \times
  10^{-1}$ (left) and  $\sigma_e = 3 \times 10^{-3}$ (right). The red
  and blue solid lines indicate $B_x$ and $B_z$ energy,
  respectively. The prediction from the linear theory is also shown by
  the dashed line.} 
  \label{weibel-in}
 \end{figure*}

 We now compare our simulation results with the maximum linear growth
 rate. Our linear analysis indicates that the electromagnetic fields
 perpendicular to the ambient magnetic field are induced by the
 electromagnetic instabilities. In fact, previous simulations showed
 that the instabilities in the shock-transition region excite $B_x$ as
 well as $B_z$ in the in-plane configuration 
 \citep[e.g.,][]{Winske1988, Matsukiyo2006}. The maximum 
 values of the $x$ and $z$ components of the magnetic field energy
 averaged over $y$ axis are determined for each snapshot, and are shown
 in Figure \ref{weibel-in} for
 $\sigma_e = 3 \times 10^{-1}$ (left) and  $\sigma_e = 3 \times 10^{-3}$
 (right). The red and blue solid line indicate the $x$ and $z$
 component of the magnetic field energy, respectively. The magnetic field
 energy $\epsilon_B = B^2/8\pi N_1m_ec^2$ are expressed in units of the
 upstream kinetic energy. The maximum linear growth rate determined by
 Equation \ref{eq:growth1} and Equation \ref{eq:growth2} is also shown
 in Figure \ref{weibel-in} with the black dashed lines. 

 For $\sigma_e = 3 \times 10^{-1}$, although the maximum linear growth
 rate is consistent with the simulation result, it is difficult
 to differentiate which instabilities generate the magnetic field
 fluctuations because the maximum growth rates (Equation
 \ref{eq:growth1} and \ref{eq:growth2}) are almost the
 same. Furthermore, our analysis assumes a cold ring distribution and
 ignores possible kinetic effects 
 that should become important at relatively short wavelength. In any
 case, we think that the instabilities excited in the shock-transition
 region generate the magnetic field fluctuations in our simulation.

 For $\sigma_e = 3 \times 10^{-3}$, the maximum linear growth rate
 gives good agreement with the simulation result. In addition, the
 maximum energy of the fluctuating magnetic field saturates about
 $10\%$--$20\%$ of the upstream bulk kinetic energy. This result is
 consistent with the previous studies
 \citep{Kato2007, Chang2008, Sironi2011}. Therefore, we conclude that
 the fluctuations in $B_x$ and $B_z$ in the shock-transition region
 result from the WI.  

 \section{Particle Energy Spectra}\label{sec:energy}

\begin{figure*}[htb!]
   \plottwo{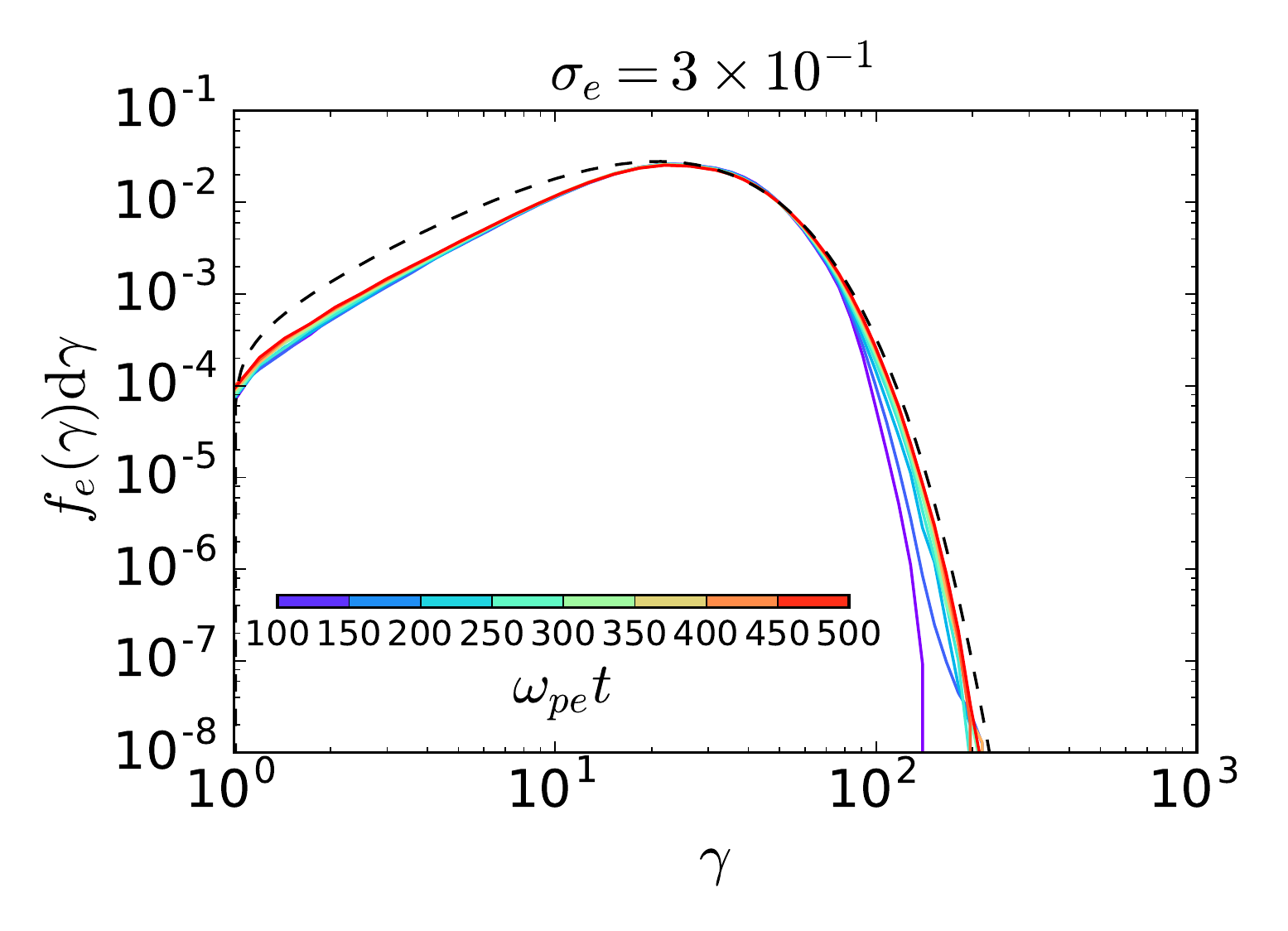}{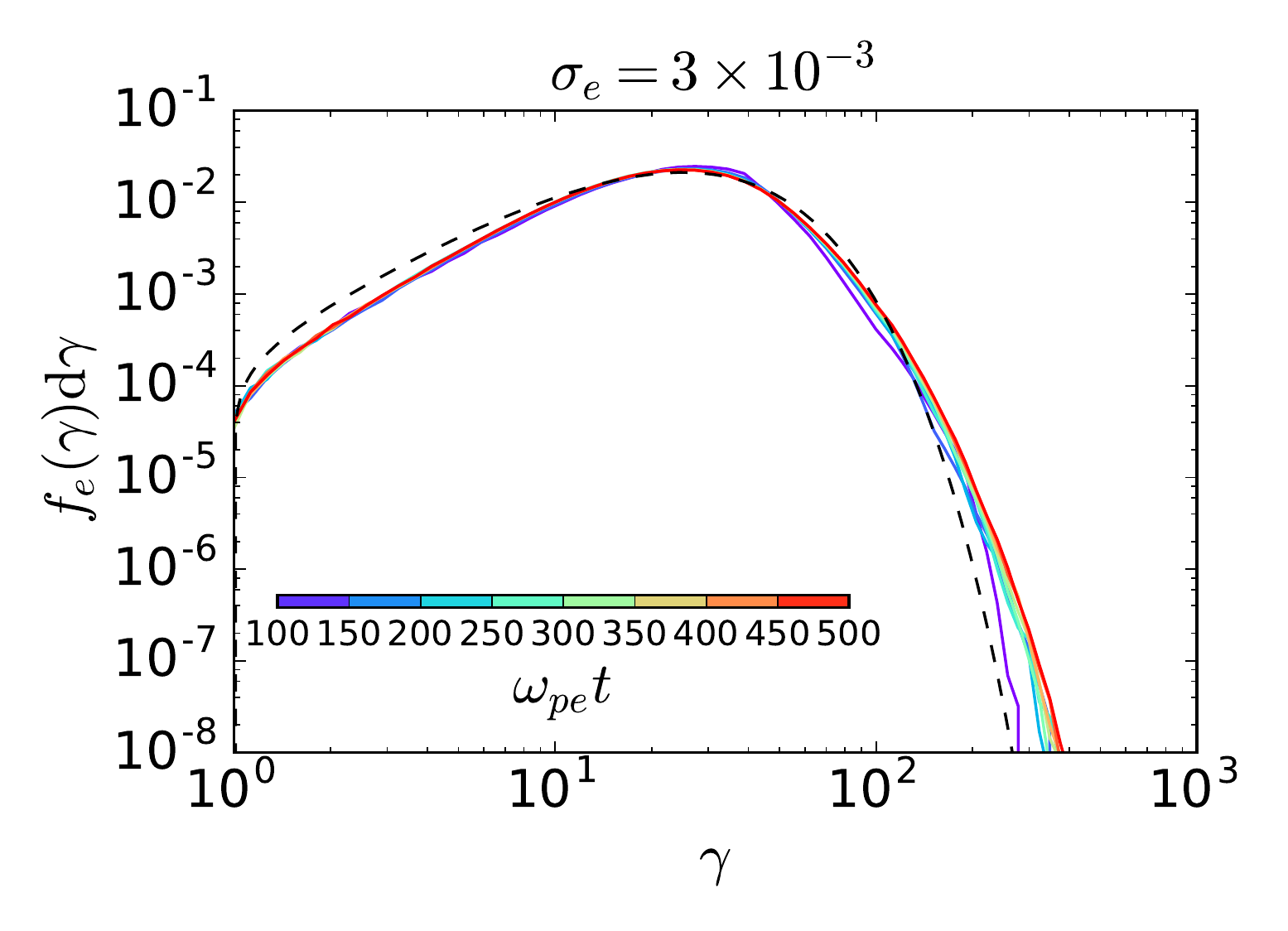}
   \caption{Downstream energy spectra of electrons:
   $\sigma_e = 3 \times 10^{-1}$ (left) and
   $\sigma_e = 3 \times 10^{-3}$ (right). The black dashed lines 
   indicate a 3D relativistic Maxwellian fitting result. }
  \label{distribution}
  \end{figure*}

  Figure \ref{distribution} shows the downstream energy spectra of
  electrons for $\sigma_e = 3 \times 10^{-1}$ and $3 \times 10^{-3}$,
  which are normalized as follows: 
  \begin{equation}
    \int^{\infty}_1 f_e({\gamma}) {\rm d}\gamma = 1.
  \end{equation} 
  The energy spectra of positrons are identical to those of
  electrons. We followed the time evolution from $\omega_{pe}t = 100$ up
  to $\omega_{pe}t = 500$. For both $\sigma_e$, the
  measured distribution reaches a steady state by the end of our
  simulation. The energy spectra can be
  well-fitted with 3D relativistic Maxwellian,
  \begin{equation}
   f(\gamma){\rm d}\gamma \propto \gamma
    \sqrt{\gamma^2-1}\exp(-\frac{\gamma m c^2}{kT}). 
  \end{equation}
  Note that the degree of freedom is three in the in-plane configuration. 
  The fitting result indicates that the downstream particles are completely
  thermalized. A clear suprathermal tail is not observed in the
  range of $\sigma_e$ used in our simulations.

  In the out-of-plane configuration, however, a suprathermal tail is
  visible for $\sigma_e = 3 \times 10^{-3}$ \citep[see][Figure
  10]{Iwamoto2017}. This agrees with the simulation result by
  \cite{Sironi2013}. They suggested that the particle 
  acceleration can be explained in terms of a Fermi process due to the
  strong turbulence generated by the WI and that the suppression of the
  cross-field diffusion \citep{Jokipii1993, Jones1998} may result in
  low-level injection of particles into the Fermi process. Our result
  also confirms that the orientation of the ambient magnetic field
  affects the efficiency of particle acceleration.

 % \bibliographystyle{aasjournal}
 % \bibliography{reference}

\end{document}